\documentclass[prd,twocolumn,nofootinbib,showpacs,amsmath,amssymb,floatfix,eqsecnum]{revtex4-1}

\usepackage{graphicx,color,framed}
\usepackage{hyperref}
\usepackage{times}
\usepackage{enumerate}
\usepackage{lipsum}
\usepackage{slashed}
\usepackage{url}

\hypersetup{
    colorlinks=true, %set true if you want colored links
    linktoc=all,     %set to all if you want both sections and subsections linked
    linkcolor=blue,  %choose some color if you want links to stand out
}

\def \beq {\begin{equation}}
\def \eeq {\end{equation}}
\def \beqa {\begin{eqnarray}}
\def \eeqa {\end{eqnarray}}
\def \bseq {\begin{subequations}}
\def \eseq {\end{subequations}}

\newcommand \ran {\rangle}
\newcommand \lan {\langle}

\newcommand \ep {\epsilon}

\newcommand \pd {\partial}

\newcommand \ov {\overline}
\newcommand \td {\tilde}
\newcommand \ZT {\mathbb{Z}_2^T}
\newcommand \T {\mathcal{T}}
\newcommand \C {\mathcal{C}}

\newcommand \nti {\eta_{\text{{\tiny{ETI}}}}}

\begin{document}

\title{Anomaly indicators for topological orders with $U(1)$ and time-reversal symmetry}

\author{Matthew F. Lapa}
\email[email address: ]{mlapa@uchicago.edu}
\affiliation{Kadanoff Center for Theoretical Physics, University of Chicago, Chicago, IL, 60637, USA}

\author{Michael Levin}
\email[email address: ]{malevin@uchicago.edu}
\affiliation{Kadanoff Center for Theoretical Physics, University of Chicago, Chicago, IL, 60637, USA}

%\date{\today}

\begin{abstract}

We study anomalies in time-reversal ($\ZT$) and $U(1)$ symmetric topological orders. In this context, an \emph{anomalous} topological order is one that cannot be realized in a strictly $(2+1)$-D system but can be realized on the surface of a $(3+1)$-D symmetry-protected topological (SPT) phase. To detect these anomalies we propose several \emph{anomaly indicators} --- functions that take as input the algebraic data of a symmetric topological order and that output a number indicating the presence or absence of an anomaly. We construct such indicators for both structures of the full symmetry group, i.e. $U(1)\rtimes\ZT$ and $U(1)\times\ZT$, and for both bosonic and fermionic topological orders. In all cases we conjecture that our indicators are complete in the sense that the anomalies they detect are in one-to-one correspondence with the known classification of $(3+1)$-D SPT phases with the same symmetry. We also show that one of our indicators for bosonic topological orders has a mathematical interpretation as a partition function for the bulk $(3+1)$-D SPT phase on a particular manifold and in the presence of a particular background gauge field for the $U(1)$ symmetry.

\end{abstract}


\maketitle

\section{Introduction}

Suppose one is given a $(2+1)$-D topological order with a global symmetry group $G$. A basic question is whether this topological order can be realized in a strictly $(2+1)$-D system. If no such realization exists, it is tempting to conclude that the topological order is unphysical. However, this is not necessarily true: some topological orders with symmetry have the property that they cannot be realized in $(2+1)$-D, but \emph{can} be realized on the boundary of certain $(3+1)$-D systems, namely $(3+1)$-D symmetry-protected topological (SPT) phases~\cite{VS2013,MKF1,MKF2,CFV-2014,wang-potter-senthil-2,BNQ,PhysRevX.5.041013,
wang-lin-levin2016}. Topological orders of this kind are said to be \emph{anomalous}. 

It is clearly desirable to be able to detect which topological orders with symmetry are anomalous and which can be realized in $(2+1)$-D. An efficient way to do this is to construct an explicit function, called an \emph{anomaly indicator}, that takes as input a topological order with symmetry and outputs a number indicating the presence or absence of an anomaly. The goal of this paper is to construct such anomaly indicators in the physically important case where the global symmetries are time-reversal and $U(1)$ symmetry.

Before proceeding, let us be more precise about our terminology. When we say ``topological order", we mean a collection of algebraic data that describes both the topological properties and symmetry quantum numbers of anyons. This data has three components: (i) a set $\mathcal{C}= \{a,b,c,\dots\}$ of anyons (or quasiparticles); (ii) fusion and braiding data for these anyons~\cite{kitaevhoneycomb}; and (iii) data that characterizes the action of the symmetry group $G$ on $\C$ 
(for general references on topological orders with symmetry, see \cite{barkeshli1,teo2015theory,tarantino2016symmetry,lan-kong-wen2}). This collection of data cannot be chosen arbitrarily: it must be physically consistent in the sense that it can be realized on the boundary of a $(3+1)$-D SPT phase with the same symmetry group $G$.\footnote{In this paper, we only consider topological orders that can be realized on surfaces of $(3+1)$-D SPT phases, which by definition are \emph{short-range entangled}. We do not address the question of which topological orders can be realized on the boundary of $(3+1)$-D long-range entangled phases. Readers interested in this topic can consult Ref.~\onlinecite{fidkowski-vishwanath2017}, for example.} The question we ask is whether a given topological order $\mathcal{C}$ can be realized on the boundary of a \emph{trivial} SPT phase (which means it can be realized in strictly $(2+1)$-D) or only on the boundary of a \emph{nontrivial} SPT phase. Anomaly indicators provide the answer to this question (and more).

Recently, Refs.~\onlinecite{levin-wang, tachikawa2017, barkeshli2} constructed anomaly indicators for topological orders whose only symmetry is time-reversal, i.e. $G = \ZT$. In this paper we build on this work by constructing anomaly indicators for topological orders with both time-reversal \emph{and} $U(1)$ symmetry. Specifically, we consider the two symmetry groups, $G = U(1)\rtimes\ZT$ and $G = U(1)\times\ZT$; in the first case, the $U(1)$ symmetry can be thought of as charge conservation, while in the second case, the $U(1)$ can be thought of as spin ($S^z$) conservation. We emphasize that the first case, $U(1)\rtimes\ZT$, is particularly important as this is the symmetry group of the electron topological insulator (ETI)~\cite{fu-kane-mele,QHZ} --- one of the few experimentally realized SPT phases~\cite{hsieh2008topological} (see also the review~\cite{hasan-kane}).

In addition to the two symmetry groups, we also consider both bosonic and fermionic\footnote{By ``fermionic topological order", we mean a topological order containing a trivial fermionic quasiparticle $e$ (representing the electron) which has trivial mutual statistics with all other quasiparticles (see Appendix~\ref{app:TO-with-e}).} topological orders. In all four cases we test our anomaly indicators on a set of nontrivial examples, and we conjecture that our indicators are \emph{complete} in the following sense: for each topological order $\C$, they completely determine which $(3+1)$-D SPT phase of a given symmetry can host $\C$ on its boundary.

We start with the case of $U(1)\rtimes\ZT$ symmetry. For bosonic topological orders with this symmetry, we propose three indicators $\eta_1$, $\eta_2$, and $\eta_3$, of which $\eta_1$ and $\eta_2$ were previously known~\cite{levin-wang,barkeshli2}. Each indicator can take on the two values $+1$ or $-1$, with a value of $-1$ indicating an anomaly. These eight possibilities are in one-to-one correspondence with the eight different phases in the $(\mathbb{Z}_2)^3$ classification of bosonic SPT phases with $U(1)\rtimes\ZT$ symmetry in $(3+1)$-D~\cite{VS2013,MKF1}. Next, for fermionic topological orders with $U(1)\rtimes\ZT$ symmetry we propose three indicators $\eta_{2,f}$, $\eta_{3,f}$, and $\nti$, of which $\eta_{2,f}$ was previously known~\cite{levin-wang,tachikawa2017}. These three indicators also take on the values $1$ or $-1$, leading to eight possibilities which correspond to the $(\mathbb{Z}_2)^3$ classification of fermionic SPT phases with $U(1)\rtimes\ZT$ symmetry in $(3+1)$-D~\cite{wang-potter-senthil,freed-hopkins}.

We then consider the case of $U(1)\times\ZT$ symmetry. For bosonic topological orders with this symmetry we propose four indicators $\eta_1$, $\eta_2$, $\eta_3$, and $\eta_4$, where the first three indicators are the same as in the $U(1)\rtimes\ZT$ case. The fourth indicator $\eta_4$ also takes the values $1$ or $-1$ only, and these four indicators combined correspond to the $(\mathbb{Z}_2)^4$ classification of bosonic SPT phases with $U(1)\times\ZT$ symmetry in $(3+1)$-D~\cite{VS2013,kapustin1}. Finally, for fermionic topological orders with $U(1)\times\ZT$ symmetry we propose the two indicators $\eta_{2,f}$ and $\eta_{3,f}$ (which already appeared in the $U(1)\rtimes\ZT$ case). In this case the indicator $\eta_{2,f}$ takes on the values $e^{i\frac{2\pi \nu}{8}}$, $\nu=0,\dots,7$, and $\eta_{3,f}$ takes on the values $1$ and $-1$. These possible values correspond to the $\mathbb{Z}_8\times\mathbb{Z}_2$ classification of fermionic SPT phases with $U(1)\times\ZT$ symmetry in $(3+1)$-D~\cite{wang-senthil,freed-hopkins}.

We give concrete physical derivations of all four anomaly indicators introduced in this paper: $\{\eta_3, \eta_4, \eta_{3,f}, \nti\}$. We also present a mathematical interpretation of the indicator $\eta_3$ as being equal to the partition function for the bulk $(3+1)$-D SPT phase on a particular manifold \emph{and} in the presence of a background $U(1)$ gauge field (see Sec.~\ref{sec:eta-3-pf} for more details). This partition function interpretation of $\eta_3$ is similar to the partition function interpretation of the indicators $\eta_1$ and $\eta_2$ from Ref.~\onlinecite{barkeshli2}.

Before closing this section, we should mention that the anomalies that we consider here are closely related to \emph{`t Hooft} anomalies in high-energy physics~\cite{hooft1980naturalness}. A quantum field theory with global symmetry of a group $G$ is said to have a `t Hooft anomaly if it cannot be consistently coupled to a background (i.e., non-dynamical) gauge field $A$ for the group $G$. What happens is that the theory cannot be regularized in such a way that the partition function $\mathcal{Z}[A]$ for the theory coupled to the background gauge field retains the full symmetry of the classical action $S[\Phi,A]$ for the theory coupled to $A$, where $\Phi$ denotes the matter fields of the theory (these are integrated out in the calculation of $\mathcal{Z}[A]\sim \int [D\Phi]e^{iS[\Phi,A]}$). It was recently understood that a powerful way to characterize an SPT phase is by the `t Hooft anomalies in theories that can appear on the boundary of that SPT phase~\cite{wen2013classifying,kapustin2014symmetry,kapustin2014anomalies,kapustin2014anomalous,
PhysRevLett.114.031601,PhysRevB.91.205101}. Although our anomaly indicators are designed to work at the level of the algebraic theory of anyons, we expect that the anomalies that they detect would also appear as `t Hooft anomalies in a continuum topological field theory description of the topological orders that we study.

This paper is organized as follows. In Sec.~\ref{sec:U1-rtimes-T} we construct anomaly indicators for bosonic and fermionic topological orders with symmetry group  $U(1)\rtimes\ZT$. In Sec.~\ref{sec:bulk-bdy} we prove various universal properties of topological insulator surfaces which we use in deriving our anomaly indicators. In Sec.~\ref{sec:U1-times-T} we construct anomaly indicators for bosonic and fermionic topological orders with symmetry group $U(1)\times\ZT$. Section~\ref{sec:conclusion} presents our conclusions. We discuss technical details in a series of four appendices. 

\section{Anomaly indicators for systems with $U(1)\rtimes\mathbb{Z}_2^T$ symmetry}
\label{sec:U1-rtimes-T}

\subsection{Bosonic case}
\label{sec:U1-rtimes-T-bosons}

\subsubsection{Review of anomaly indicators for systems with $\mathbb{Z}_2^T$ symmetry}

We begin by reviewing the anomaly indicators for time-reversal symmetric bosonic topological 
orders~\cite{levin-wang, barkeshli2}.

To begin, recall that in $(3+1)$-D, there are believed to be four distinct bosonic SPT phases with time-reversal symmetry, which are classified by the group $\mathbb{Z}_2 \times \mathbb{Z}_2$~\cite{VS2013}. It follows from this result that every time-reversal symmetric bosonic topological order is associated with a $\mathbb{Z}_2 \times \mathbb{Z}_2$ valued anomaly. Ref.~\cite{levin-wang} proposed that this anomaly can be diagnosed by two indicators $\eta_1$ and $\eta_2$, which can each take on the values $\pm 1$ with a value of $-1$ indicating an anomaly. The first indicator $\eta_1$ takes the form
\beq
	\eta_1= \frac{1}{D}\sum_{a\in\C} d_a^2 e^{i\theta_a}\ ,
\label{eta1}
\eeq 
where $\mathcal{C}$ denotes the set of anyons in the topological order. Here, $d_a$ (a positive real number) is the quantum dimension of the anyon $a$, while $D=\sqrt{\sum_a d_a^2}$ is the total quantum dimension, and $\theta_a$ is the topological spin of the anyon $a$ (e.g., $\theta_a=\pi$ for a fermion). 

The second indicator $\eta_2$ takes the form
\beq
	\eta_2= \frac{1}{D}\sum_{a\in\C} d_a \T^2_a e^{i\theta_a}\ ,
\label{eta2}
\eeq
where the quantity $\T^2_a$ is defined as follows. First, let $\T(a)$ denote the image of the anyon $a$ under the action of time-reversal. We always have $\T(\T(a))= a$ since time-reversal is an involution. Anyons which are invariant under time-reversal, $\T(a)=a$, have two options as to how the time-reversal symmetry is implemented on them: they can either be local Kramers singlets or local Kramers doublets (see  Ref.~\onlinecite{levin-stern} for a precise definition of local Kramers degeneracy). Then $\T^2_a$ is defined as
\begin{align*}
	\T^2_a= \begin{cases}
	0\ & ,\ \T(a) \neq a \\
	1\ & ,\ \T(a)=a\text{ and $a$ is a Kramers singlet} \\
	-1\ & ,\ \T(a)=a\text{ and $a$ is a Kramers doublet}
	\end{cases}
\end{align*}

What is the physical interpretation of $\eta_1$ and $\eta_2$? We can get some insight into the first indicator $\eta_1$ by recalling that for any strictly two-dimensional bosonic topological order, $\eta_1 = e^{i\frac{2\pi}{8}c_{-}}$ where $c_-$ is the chiral central charge of the edge modes at the boundary~\cite{kitaevhoneycomb}. This relation gives a simple proof that $\eta_1 = 1$ for strictly two-dimensional time-reversal symmetric topological orders, since $c_-$ is odd under time-reversal.

While we do not know of a similarly concrete picture for $\eta_2$, there is a field-theoretic interpretation of $\eta_2$ (as well as $\eta_1$) as the partition function $\mathcal{Z}(X)$ of the bulk $(3+1)$-D SPT phase on an appropriate closed Euclidean four-manifold $X$. In particular, $\eta_1 = \mathcal{Z}(\mathbb{CP}^2)$, the partition function on the (orientable) Euclidean four-manifold $\mathbb{CP}^2$, while $\eta_2 = \mathcal{Z}(\mathbb{RP}^4)$, the partition function on the (non-orientable) Euclidean four-manifold $\mathbb{RP}^4$~\cite{barkeshli2}. 

\subsubsection{An additional indicator $\eta_3$}

Having warmed up with case of time-reversal symmetry, we now move on to discuss systems with \emph{both} charge conservation and time-reversal symmetry, i.e.
$U(1)\rtimes \ZT$ symmetry. In this case, there are believed to be $8$ distinct SPT phases in $(3+1)$-D with a $(\mathbb{Z}_2)^3$ classification~\cite{VS2013,MKF1} (one $\mathbb{Z}_2$ factor is beyond the group cohomology classification, which yields only a $(\mathbb{Z}_2)^2$ classification~\cite{CGLW2013}). Of these three $\mathbb{Z}_2$ factors, the first two are generated by bosonic SPTs with only $\ZT$ symmetry, and so the $U(1)$ symmetry plays no role in those phases. In particular, this means that the $\eta_1$ and $\eta_2$ indicators discussed above are sufficient for detecting the associated anomalies. As for the third $\mathbb{Z}_2$ factor, this is generated by the bosonic topological insulator (BTI) phase~\cite{VS2013,MKF1}. Thus, our task is to find an anomaly indicator, $\eta_3$, that can detect the $\mathbb{Z}_2$ anomaly associated with the BTI phase.

We propose that the missing indicator takes the form
\beq
	\eta_3= \frac{1}{D}\sum_{a\in\C} d_a^2  e^{i\theta_a}  e^{i2\pi q_a} \,
\label{Eta3}
\eeq
where $q_a$ denotes the $U(1)$ \emph{charge} of the anyon $a$. Here we use a convention where the fundamental (local) bosons in the system have charge $1$ so that $q_a$ is determined modulo $1$. 

In the remainder of this section, we will show that $\eta_3$ has all the properties that we want in an anomaly indicator:
\begin{enumerate}
\item{$\eta_3$ can only take the values $+1$ and $-1$.}

\item{Any topological order that can be realized in strictly two dimensions has $\eta_3 = +1$.}

\item{Any topological order that exists on the boundary of the bosonic topological insulator has $\eta_3 = -1$.}

\end{enumerate}
(Here all topological orders are $U(1)\rtimes\ZT$ symmetric).

\subsubsection{Examples}

In this section we compute the values of the indicators $\eta_1$, $\eta_2$, and $\eta_3$ for three important examples, namely the ``EfMf'', ``ETMT'', and ``ECMC'' topological orders. These topological orders are interesting because they are believed to appear on the surfaces of the three `root' phases in the $(\mathbb{Z}_2)^3$ classification of $(3+1)$-D bosonic SPTs with $U(1)\rtimes\ZT$ symmetry~\cite{VS2013}.  

Let us recall the definitions of these three topological orders. The ETMT and ECMC topological orders have the same anyons and same statistics as the usual toric code model, i.e. $\C=\{1,E,M,\ep\}$ where $E$ and $M$ are bosons and $\ep = E \times M$ is a fermion. What makes these topological orders interesting is their symmetry assignments. In particular, in the ETMT theory, all anyons carry charge zero, but the $E$ and $M$ particles are Kramers doublets, and the $1$ and $\ep$ particles are Kramers singlets. Likewise, in the ECMC theory, all anyons are Kramers singlets, but the $E$ and $M$ particles carry charge $1/2$, and the $1$ and $\ep$ particles carry charge zero. The EfMf topological order is slightly different from the other two examples: this state has four anyons $\C=\{1,E,M,\ep\}$ with the same fusion rules as the toric code, but $E, M, \ep$ are \emph{fermions}. The symmetry assignments in EfMf are all trivial: all the anyons are Kramers singlets with charge zero. 

In Table~\ref{tab:U1-rtimes-T-bosons}, we list the values of the indicators $\eta_1, \eta_2, \eta_3$ for the three bosonic topological orders discussed above. We can draw several conclusions from this table. First, we can see that $\eta_1, \eta_2, \eta_3$ are \emph{independent}, in the sense that the values in the table are inconsistent with the possibility that any one indicator might be equal to a product of some of the others. Second, we can see that the three indicators $\eta_1$, $\eta_2$, and $\eta_3$ succeed in detecting the anomalies in the EfMf, ETMT, and ECMC topological orders, as well as in any combination of these topological orders, e.g. ETMT $\times$ ECMC, etc. The latter result is particularly significant since, as we mentioned earlier, these three topological orders are believed to describe the surfaces of the three `root' phases with $U(1)\rtimes\ZT$ symmetry. Assuming this is the case, it follows that $\eta_1$, $\eta_2$, and $\eta_3$ provide a \emph{complete} set of anomaly indicators for bosonic topological orders with $U(1)\rtimes\ZT$ symmetry.

\begin{table}[t]
\begin{center}
\begin{tabular}{ c |c| c| c}
  & EfMf & ETMT & ECMC \\ \hline
$\eta_1$ & $-1$ & 1 & 1 \\ \hline
$\eta_2$ & $-1$ & $-1$ & 1\\ \hline
$\eta_3$ & $-1$ & 1 & $-1$ \\ 
\end{tabular} 
\caption{\label{tab:U1-rtimes-T-bosons} Values of the indicators $\eta_1$, $\eta_2$, $\eta_3$ for the EfMf, ETMT, and ECMC topological orders which can appear on the surfaces of the three root phases in the 
$(\mathbb{Z}_2)^3$ classification of bosonic SPTs with $U(1)\rtimes\ZT$ symmetry in $(3+1)$-D.}
\end{center}
\end{table}

\subsubsection{Physical picture and derivation of the indicator $\eta_3$}
\label{sec:boson-physical-picture}

We now present a derivation of the indicator $\eta_3$ and its properties. Our derivation is based on three results about charge conserving and time-reversal symmetric bosonic topological orders. The first result (proven below) is that every $U(1)$ symmetric bosonic topological order $\C$ contains a unique Abelian anyon $f$ with the property that the mutual statistics between $f$ and any other anyon $a$ is given by
\begin{align}
e^{i\theta_{a,f}} = e^{i2\pi q_a} \label{eq:AB-phase}
\end{align}
[By mutual statistics, we mean the statistical phase for a process in which the anyon $a$ makes a complete circuit around the anyon $f$]. We refer to $f$ as the ``$2\pi$ flux anyon'', or just the ``flux anyon''. The second result, which we prove in Sec.~\ref{sec:bulk-bdy}, is that any $U(1)\rtimes\ZT$ symmetric bosonic topological order $\C$ that can appear on the surface of the BTI must have the property that $f$ is a \emph{fermion}:
\begin{align}
e^{i \theta_f} = -1
\end{align}
The third result, also proven in Sec.~\ref{sec:bulk-bdy}, is that any $U(1)\rtimes\ZT$ symmetric bosonic topological order $\C$ that can be realized in strictly two dimensions must have the property that $f$ is a \emph{boson}: 
$e^{i\theta_f} = +1$.

These results suggest a natural way to construct an anomaly indicator: we will define a function $\td{\eta}_3$ that takes as input a $U(1)\rtimes\ZT$ symmetric topological order $\C$, and returns as output the topological spin $e^{i\theta_f}$ of the flux anyon $f$. We are then guaranteed that $\td{\eta}_3 = -1$ for any topological order supported on the surface of the BTI, while $\td{\eta}_3 = +1$ for any strictly two-dimensional topological order. 

Following this plan, we now proceed to construct the function $\td{\eta}_3$. Our construction begins with the following theorem, which is proved on Pg.~10 of Ref.~\onlinecite{barkeshli1}. Suppose there is a function $\phi:\C \to \mathbb{R}$, $a \mapsto \phi_a$, such that $e^{i\phi_a}e^{i\phi_b}= e^{i\phi_c}$ whenever $N^c_{ab}\neq 0$ (here, $N^c_{ab}$ is the number of ways that anyons $a$ and $b$ can fuse to form anyon $c$). Then
\begin{align*}
	e^{i\phi_a}= M^*_{af}
\end{align*}
for some unique\footnote{The uniqueness of $f$ was not explicitly stated in Ref.~\onlinecite{barkeshli1}, but follows immediately from the unitarity of the $S$-matrix.} Abelian anyon $f$, where $M_{ab}$ is defined by
\beq
	M_{ab}=\frac{S^*_{ab}S_{11}}{S_{1a}S_{1b}}\ . \label{eq:MS}
\eeq
Here $1$ denotes the identity anyon, $S$ denotes the $S$-matrix, and $^*$ denotes complex conjugation. The matrix $M_{ab}$ is known as the ``monodromy scalar component.'' When $M_{ab}$ is a $U(1)$ phase, it can be written as
\beq
	M_{ab}= e^{i\theta_{a,b}}\ ,
\eeq
where $\theta_{a,b}=\theta_{b,a}$ has the physical interpretation (see Eq.~39 of \cite{barkeshli1}) of being equal to the 
\emph{mutual statistics} for a process in which anyon $b$ makes a counterclockwise circuit around anyon $a$. Recall also that for a bosonic topological order $\C$, the $S$ matrix is unitary and symmetric, and $S_{1a}=\frac{d_a}{D}$ for any $a$. 

We now apply this theorem to topological orders with $U(1)$ symmetry by choosing the function
$\phi_a$ to be $\phi_a= -2\pi q_a$. In this case the Abelian anyon $f$ satisfies
\beq
	e^{i2\pi q_a}= M_{af}\  \label{eq:monodromy}
\eeq
and, since $M_{af}= e^{i\theta_{a,f}}$, the Abelian anyon $f$ appearing here is exactly the $2\pi$ flux anyon mentioned above. This proves our first claim: for any bosonic topological order $\C$ with $U(1)$ symmetry,
there exists a unique Abelian anyon $f$ satisfying Eq.~\eqref{eq:AB-phase}.

Next, consider the Kronecker delta function $\delta_{bf}$ on $\C$, which is equal to one if the anyon $b$ is equal to the flux anyon $f$ and equal to zero otherwise. The theorem quoted above allows us to write $\delta_{bf}$ in the form
\beq
	\delta_{bf} = \frac{1}{D}\sum_{a\in\C}d_a S_{ab} e^{i2\pi q_a} \ . \label{eq:NA-formula}
\eeq
This can be checked by plugging in for $e^{i2\pi q_a}$ from Eq.~\eqref{eq:monodromy} and using the unitarity of the $S$ 
matrix, combined with the fact that $d_f=1$ (since $f$ is Abelian by the theorem quoted above). Using this expression for $\delta_{bf}$, it is now easy to construct $\td{\eta}_3$:
\begin{align}
	\td{\eta}_3 = e^{i\theta_f} = \frac{1}{D}\sum_{a,b\in\C}d_a S_{ab} e^{i2\pi q_a} e^{i\theta_b}		
\end{align}

In principle, we could declare victory and stop here: as we mentioned above, $\td{\eta}_3$ is \emph{guaranteed} to take the value $+1$ for strictly two-dimensional topological orders, and $-1$ for topological orders that live on the surface of a bosonic topological insulator. Thus, $\td{\eta}_3$ has the essential properties that we want in an anomaly indicator. 

That said, we find it convenient to use a slightly different indicator, defined by:
\beq
	\eta_3 \equiv \td{\eta}_3/\eta_1\ .
\label{Eta32}
\eeq
Like $\td{\eta}_3$, the indicator $\eta_3$ is guaranteed to take the value $+1$ for strictly two-dimensional topological orders and $-1$ for topological orders that live on the surface of a bosonic topological insulator, since $\eta_1 = 1$ in both cases. Our motivation for using $\eta_3$ instead of $\td{\eta}_3$ is that (i) $\eta_3$ has a simpler algebraic expression than $\td{\eta}_3$ and (ii) it generalizes more easily to the fermionic case, as we will see later.

To complete our derivation of $\eta_3$, all that remains is to show that the above definition of $\eta_3$ (\ref{Eta32}) is equivalent to our previous definition (\ref{Eta3}). To prove this, we first rewrite $\td{\eta}_3$ as 
\begin{align*}
	\td{\eta}_3=  \frac{1}{D}\sum_{a,b\in\C} d_a d_b S_{ab} e^{i2\pi q_a}e^{i\theta_b}
\end{align*}
where we have inserted an extra factor of $d_b$. This does not change $\td{\eta}_3$ since the sum over $a$ gives $\delta_{bf}$ and $d_f=1$ as $f$ is Abelian. Next, we relabel summation indices $a\leftrightarrow b$ and then use $S_{ba}=S_{ab}$ to rewrite $\td{\eta}_3$ as
\begin{align*}
	\td{\eta}_3=  \frac{1}{D}\sum_{a,b\in\C} d_a d_b S_{ab} e^{i2\pi q_b}e^{i\theta_a}\ .
\end{align*}
We now insert the explicit formula $S_{ab}=\frac{1}{D}\sum_{c\in\C} N^c_{a\ov{b}} d_c e^{i\theta_c}e^{-i\theta_a}e^{-i\theta_b}$ (Eq.~37 in \cite{barkeshli1}) to find that
\begin{align*}
	\td{\eta}_3=  \frac{1}{D^2}\sum_{a,b,c\in\C} d_a d_b d_c N^c_{a\ov{b}} e^{i\theta_c}e^{-i\theta_b}  e^{i2\pi q_b}\ .
\end{align*}
Next, we use $N^c_{a\ov{b}}= N^a_{bc}$ and $\sum_{a\in \C} N^a_{bc} d_a = d_b d_c$ to find that
\begin{align}
	\td{\eta}_3 &=  \frac{1}{D^2}\sum_{b,c\in\C} d_b^2 d_c^2  e^{i\theta_c}e^{-i\theta_b}  e^{i2\pi q_b} \nonumber \\
	&= \left(\frac{1}{D} \sum_{c\in\C} d_c^2 e^{i\theta_c}\right)\left(\frac{1}{D}\sum_{b\in\C} d_b^2   
e^{-i\theta_b}  e^{i2\pi q_b} \right) \nonumber \\
	&= \eta_1 \left(\frac{1}{D}\sum_{b\in\C} d_b^2   
e^{-i\theta_b}  e^{i2\pi q_b} \right)\ . \label{eq:eta3-eqn}
\end{align}
For the final step, we change summation variables in the second factor on the right-hand side by setting $b= \T(a)$. Since $d_{\T(a)}=d_a$, $q_{\T(a)}=q_a$, and $\theta_{\T(a)}=-\theta_a$, we find that
\begin{align*}
	\td{\eta}_3 = \eta_1 \left(\frac{1}{D}\sum_{a\in\C} d_a^2   
e^{i\theta_a}  e^{i2\pi q_a} \right) = \eta_1\eta_3\ ,
\end{align*}
which completes the proof. 

\subsubsection{Proof that the indicator $\eta_3$ takes the values $\pm 1$}
To complete our analysis of $\eta_3$, we need to show that $\eta_3$ only takes the values $\pm 1$ for $U(1)\rtimes\ZT$ symmetric topological orders. We will do this in two steps. 

First, we will show that $\eta_3$ is always \emph{real}. To see this, we note that
\begin{align*}
	\eta_3^* = \frac{1}{D}\sum_{a\in\C} d_a^2  e^{-i\theta_a}  e^{-i2\pi q_a}\ .
\end{align*}
Next, we change summation variables by setting $a= \ov{b}$ ($\ov{b}$ is the unique anti-particle to $b\in\C$), and we use the three properties $q_b+q_{\ov{b}}= 0$ (mod $1$), $d_b=d_{\ov{b}}$, and $\theta_b=\theta_{\ov{b}}$ (mod $2\pi$) to find
\begin{align*}
	\eta_3^* = \frac{1}{D}\sum_{b\in\C} d_b^2  e^{-i\theta_b}  e^{i2\pi q_b}\ .
\end{align*}
Finally, we change summation variables again by setting $b=\T(c)$, and we use $q_c= q_{\T(c)}$ (mod $1$), $d_c=d_{\T(c)}$ and $\theta_c=-\theta_{\T(c)}$ (mod $2\pi$) to find that 
\begin{align*}
	\eta_3^* = \frac{1}{D}\sum_{c\in\C} d_c^2  e^{i\theta_c}  e^{i2\pi q_c} = \eta_3\ ,
\end{align*}
which proves that $\eta_3$ is a real number.

Next, we recall that $\td{\eta}_3$ is always a $U(1)$ phase (this follows from the definition). Combining these two facts with the relation $\td{\eta}_3= \eta_1 \eta_3$, it is easy to see that $\eta_3$ and $\td{\eta}_3$ equal $\pm 1$ for systems with $U(1)\rtimes\ZT$ symmetry. To prove this, we take the absolute value of the equation $\td{\eta}_3= \eta_1 \eta_3$. Since $\eta_1=\pm 1$ and $\td{\eta}_3$ is a $U(1)$ phase, this gives $|\eta_3|=1$. We already know that $\eta_3$ is real, and so this implies that $\eta_3=\pm 1$. Then, since $\eta_1=\pm 1$, this implies that $\td{\eta}_3=\pm 1$ as well. Thus, we find that $\eta_3$ and $\td{\eta}_3$ can only take on the values $\pm 1$ in systems with $U(1)\rtimes\ZT$ symmetry.

\subsubsection{$\eta_3$ as a partition function on $\mathbb{CP}^2$}
\label{sec:eta-3-pf}

We now show that just like $\eta_1$ and $\eta_2$, the indicator $\eta_3$ can be interpreted as a partition function for a $(3+1)$-D bosonic SPT phase on a certain Euclidean four-manifold. The additional element that enters in the case of $\eta_3$ is that it turns out to be a partition function on a Euclidean four-manifold \emph{and} in the presence of a nontrivial background electromagnetic field (this field probes the $U(1)$ symmetry of the SPT phase). 

To demonstrate that $\eta_3$ is a partition function we first recall the work of Kapustin~\cite{kapustin1}, who considered all possible forms for the effective action of a (3+1)-D bosonic SPT phase with symmetry group $U(1)\rtimes\mathbb{Z}^T_2$ on a Euclidean four-manifold $X$. This effective action uses topological data associated with the manifold $X$, and it also includes the effect of coupling to an external electromagnetic field parametrized by the one-form potential $A=A_{\mu}dx^{\mu}$. When the external field $A$ is set to zero, the partition function for a SPT phase on $X$ (which is the exponential of the effective action) can be written in the form (Eq.~196 of \cite{barkeshli2})
\beq
	\mathcal{Z}(X)= e^{i n_1 \pi \int_X w_2\wedge w_2}e^{i n_2\pi \int_X w_4}\ .
\eeq
Here, $w_2$ and $w_4$ are the second and fourth \emph{Stiefel-Whitney}{\footnote{Here
we use a less precise notation which is common in the recent physics literature. The mathematically correct
notation for the integral expression $\int_X w_2\wedge w_2$  is $\lan w_2\cup w_2,[X]\ran$, where $\cup$ is the cup product in 
cohomology, $[X] \in H_4(X,\mathbb{Z}_2)$ is the fundamental class of $X$ (the 4-cycle in homology which represents 
the entire manifold $X$), and $\lan \cdot, \cdot\ran$ is the pairing between elements of the homology and cohomology groups. See Ref.~\onlinecite{milnor-stasheff} for details.}} classes of $X$, and $n_1$ and $n_2$ are integers taking the values $\{0,1\}$. These possible values for $n_1$ and $n_2$ correspond to the $(\mathbb{Z}_2)^2$ group structure of $(3+1)$-D bosonic SPT phases with $\ZT$ symmetry only, and each SPT phase corresponds to a particular choice of $n_1$ and $n_2$.

Next, Kapustin showed that when the external field $A$ is turned on, the only new term that needs to be included in the effective action is the theta term 
\beq
	S_{\Theta}[A]= \frac{\Theta}{8\pi^2}\int_X F\wedge F \label{eq:theta-term}
\eeq
which can have $\Theta=0,2\pi$ (mod $4\pi$) for bosonic SPT phases with $U(1)\rtimes\ZT$ symmetry ~\cite{VS2013}. Then the full partition function in the presence of the external field $A$ is 
\beq
	\mathcal{Z}_A(X)=\mathcal{Z}(X)e^{i \frac{\Theta}{8\pi^2}\int_X F\wedge F}\ .
\eeq
Note also that the set of numbers $n_1=\{0,1\}$, $n_2=\{0,1\}$, and $\Theta=\{0,2\pi\}$ specifies the particular bosonic SPT phase with $U(1)\rtimes \mathbb{Z}^T_2$ symmetry in the $(\mathbb{Z}_2)^3$ classification of these phases.

With this background, we can now derive our partition function interpretation of $\eta_3$. First, consider the four-manifold $X=\mathbb{CP}^2$. In addition, consider a particular choice $A=A_{2\pi}$ of the external field, where $A_{2\pi}$ is chosen so that the field strength $F_{2\pi}=dA_{2\pi}$ has $2\pi$ flux through the single nontrivial 2-cycle $\mathcal{M}_2$ of $\mathbb{CP}^2$, $\int_{\mathcal{M}_2} \frac{F_{2\pi}}{2\pi}=1$. One concrete way to realize this is to choose $F_{2\pi}=2K$, where $K$ is the \emph{Kahler two-form} associated with the Fubini-Study metric on $\mathbb{CP}^2$ (see Appendix B of \cite{lapa2017} for details). In this case one can show that
\beq
	\frac{\Theta}{8\pi^2}\int_{\mathbb{CP}^2} F_{2\pi}\wedge F_{2\pi} = \frac{\Theta}{2}\ .
\eeq
Combining this result with the relation~\cite{barkeshli2} 
$\mathcal{Z}(\mathbb{CP}^2) = \eta_1$, we conclude that
\beq
	\mathcal{Z}_{A_{2\pi}}(\mathbb{CP}^2)= \eta_1 e^{i\frac{\Theta}{2}}\ .
\eeq	

To finish the process of writing $\eta_3$ as a partition function, we now relate $\Theta$ to $\theta_f$ and therefore to $\td{\eta}_3$. Suppose we have a SPT phase such that the effective action in the bulk of this phase includes the theta term $S_{\Theta}[A]$. Then if $X$ has a boundary this action reduces (locally) to $\frac{\Theta}{2\pi}\frac{1}{4\pi}\int_{\pd X} A\wedge F$, indicating a boundary quantum Hall response with Hall conductivity $\sigma_H=\frac{\Theta}{2\pi}$. Now, we show in Sec.~\ref{sec:bulk-bdy} that there is a relation between the Hall conductivity of short-range entangled $U(1)$ conserving boundaries, and the statistics of the $2\pi$ flux anyon $f$ in $U(1)\rtimes\ZT$ symmetric topological orders: $e^{i \theta_f} = e^{i \pi \sigma_H}$. This allows us to make the identification
\beq
	\td{\eta}_3 = e^{i\frac{\Theta}{2}}\ ,
\eeq
since $\td{\eta}_3 = e^{i\theta_f}$ by construction. Finally, since we have $\td{\eta}_3=\eta_1\eta_3$ and $\eta_3=\eta_1\td{\eta}_3$ (since all indicators here square to $1$), we find that
\beq
	\eta_3= \mathcal{Z}_{A_{2\pi}}(\mathbb{CP}^2)\ .
\eeq

Therefore we have shown that $\eta_3$, evaluated on a topological order $\C$ that appears on the boundary of a bosonic SPT phase, is equal to the partition function of that bulk SPT phase on the closed four-manifold $\mathbb{CP}^2$ \emph{and} in the presence of a particular configuration $A_{2\pi}$ of the external electromagnetic field $A$ on $\mathbb{CP}^2$.

\subsection{Fermionic case}
\label{sec:U1-rtimes-T-fermions}

We now consider the case of fermionic systems with $U(1)\rtimes\ZT$ symmetry\footnote{Strictly speaking, the symmetry group of the systems we consider is not quite $U(1)\rtimes\ZT$ since $\T^2 = P_F$: instead, $U(1)\rtimes\ZT = G_{\text{full}}/\{1,P_F\}$ where $G_{\text{full}}$ is the full symmetry group. See Sec.~3.5 of Ref.~\onlinecite{guo2018time} for further discussion on this point.
There, the authors also use an alternative notation ``$\frac{U(1)\rtimes \mathbb{Z}^T_4}{\mathbb{Z}_2}$''
for the group that we call $G_{\text{full}}$.} where the time-reversal operation obeys 
\begin{align*}
\T^2=(-1)^Q = P_F
\end{align*}
with $P_F$ being fermion parity and $Q$ being the charge operator, i.e. the generator of the $U(1)$ symmetry. The main objects of our study are ``fermionic topological orders", i.e. topological orders $\C$ featuring a trivial fermionic quasiparticle $e$, which we refer to as the electron.\footnote{The electron $e$ should not be confused with the electric particle in the toric code, which we denote by $E$.} Here, triviality of $e$ means that it has trivial mutual statistics with every quasiparticle $a\in\C$, i.e. $\theta_{e,a}= 0$ (mod $2\pi$). In addition, $e$ has charge $q_e=1$ and is a local Kramers doublet under time-reversal, $\T^2_e=-1$, as is appropriate for a physical electron. We review the properties of these fermionic topological orders in Appendix~\ref{app:TO-with-e}, and we recommend that readers glance at that appendix before reading the rest of this section.

\subsubsection{Review of anomaly indicators for systems with $\ZT$ symmetry}
\label{sec:ferm-review}

We begin by reviewing the known anomaly indicators for time-reversal symmetric fermionic systems~\cite{levin-wang,tachikawa2017}. In $(3+1)$-D there are believed to be 16 time-reversal symmetric fermionic SPT phases, corresponding to the $\mathbb{Z}_{16}$ classification of interacting $(3+1)$-D topological superconductors in symmetry class DIII~\cite{fidkowski-TSC,metlitski2014,wang-senthil,kitaevZ16,morimoto2015breakdown,kapustin2015fermionic,witten2016,
witten-parity,seiberg2016gapped,Hsieh2016,tachikawa2016gauge,freed-hopkins,tachikawa-yonekura-TR} (note that the non-interacting classification is $\mathbb{Z}$ instead of $\mathbb{Z}_{16}$~\cite{schnyder2008,kitaev2009periodic}). It follows from this result that every time-reversal symmetric fermionic topological order is associated with an anomaly valued in $\mathbb{Z}_{16}$, which can be realized concretely by a phase $e^{i\frac{2\pi \nu}{16}}$ for $\nu=0,\dots,15$. Ref.~\onlinecite{levin-wang} proposed that this anomaly (the phase $e^{i\frac{2\pi \nu}{16}}$) could be detected by an indicator $\eta_{2,f}$ of the form
\beq
	\eta_{2,f}= \frac{1}{D\sqrt{2}}\sum_{a\in\C} \td{\T}^2_a d_a e^{i\theta_a}\ ,
\eeq
where the sum is taken over the entire fermionic topological order $\C$ (i.e., the sum includes $b$ and $b\times e$ for any anyon $b$), $D$ is the total quantum dimension of the entire topological order including the electron, and $\td{\T}^2_a$ is a new quantity which we define below. It was conjectured in Ref.~\onlinecite{levin-wang} that $\eta_{2,f}$ equals a 16th root of unity for any fermionic topological order $\C$ with $\ZT$ symmetry and that $\eta_{2,f} \neq 1$ indicates an anomaly. A field-theoretic derivation of this indicator was later given in Ref.~\onlinecite{tachikawa2017}, which was based on those authors' previous study of time-reversal symmetric topological field theories on a crosscap background~\cite{tachikawa-yonekura-TR}. 

The new quantity $\td{\T}^2_a$ appearing in $\eta_{2,f}$ is defined as follows. For fermionic topological orders, it is possible for the time-reversal transformation to map an anyon into itself fused with the electron, $\T(a)=a\times e$. Anyons that transform under time-reversal in this way can be consistently assigned a value of 
$\T^2_a=\pm i$~\cite{metlitski2014}. Likewise, anyons with $\T(a) = a$ can be assigned a value $\T^2_a = \pm 1$ depending on whether they are Kramers singlets or Kramers doublets, in the same way as the bosonic case. The quantity $\td{\T}^2_a$ is then defined as
\begin{align*}
	\td{\T}^2_a = \begin{cases}
		\T^2_a & \ ,\  \T(a)=a \\
		-i\T^2_a & \ ,\ \T(a)= a\times e \\
		0 & \ ,\ \text{otherwise}\ .
	\end{cases}
\end{align*}

\subsubsection{Additional indicators $\eta_{3,f}$ and $\nti$}
\label{sec:eta3feti}
We now move on to the case of fermionic topological orders with $U(1)\rtimes\ZT$ symmetry. The full classification of interacting $(3+1)$-D fermionic SPT phases with $U(1)\rtimes\ZT$ symmetry is believed to be $(\mathbb{Z}_2)^3$~\cite{wang-potter-senthil,freed-hopkins}. One of the $\mathbb{Z}_2$ factors in this classification is generated by the ordinary electron topological insulator (ETI) phase~\cite{fu-kane-mele,QHZ}, which exists even within the non-interacting classification based on band theory. The other two $\mathbb{Z}_2$ factors exist only in the presence of interactions. As discussed in Ref.~\onlinecite{wang-potter-senthil}, the phases that generate these other $\mathbb{Z}_2$ factors can be thought of as combinations of bosonic SPT phases and trivial insulating states of electrons. For one of these two $\mathbb{Z}_2$ factors, the corresponding fermionic SPT phase requires only $\ZT$ symmetry and therefore the associated anomaly can be detected by the indicator $\eta_{2,f}$, reviewed above. For the other $\mathbb{Z}_2$ factor, the corresponding fermionic SPT phase can be thought of as a combination of a bosonic topological insulator built from \emph{charge 2 bosons}, and a trivial insulating state of electrons. We will refer to this phase as the ``BTI$+e$" phase. Our task is to find indicators that can detect the anomalies associated with the ETI and ``BTI$+e$" states.

To detect the anomaly on the surface of the BTI$+e$ state, we propose the indicator 
\beq
	\eta_{3,f}=  \frac{1}{D\sqrt{2}}\sum_{a\in\C}  d^2_a e^{i\theta_a}e^{i\pi q_a}\ ,
\label{eta3f}
\eeq
where $q_a$ again denotes the charge of the quasiparticle $a$, and the sum again runs over all quasiparticles in $\C$ including the electron $e$. Note that all charges are now defined \emph{modulo $2$} since the electron carries charge $q_e = 1$ in our convention.\footnote{This is because the electron $e$ has charge $1$, so local operators can only change the charge of an excitation modulo $2$.} This explains the factor of two difference between the coefficient of $q_a$ in $\eta_{3,f}$ and the previous bosonic indicator $\eta_3$ from Eq.~\eqref{Eta3}. 
We now argue that $\eta_{3,f}$ has all of the properties that we want:
\begin{enumerate}
\item{$\eta_{3,f}$ can only take the values $+1$ and $-1$.}

\item{Any topological order that can be realized in strictly two dimensions has $\eta_{3,f} = +1$.}

\item{Any topological order that exists on the boundary of the BTI$+e$ state has $\eta_{3,f} = -1$.}
\end{enumerate}
(As in the bosonic case, the topological orders mentioned in these properties are all $U(1)\rtimes\ZT$ symmetric).

At the present time we do not have a general proof of property $1$ but we will see that it holds in all the examples discussed below. As for properties $2$ and $3$, they both follow from the formula
\begin{align}
\eta_{3,f} = e^{i \frac{2\pi}{8} (c_- - \sigma_H)},
\label{eta3fid}
\end{align}
which holds for any strictly two-dimensional fermionic topological order with $U(1)$ symmetry only (see Appendix~\ref{app:eta3f} for a derivation). Property 2 follows immediately from (\ref{eta3fid}) by noting that time-reversal symmetry forces $c_- = \sigma_H = 0$. Likewise, to establish property $3$, recall one of the fundamental characteristics of the BTI$+e$ phase: it hosts a surface state that is $U(1)$ conserving, time-reversal breaking, and short-range entangled, with the additional feature that the $(1+1)$-D edge between this surface state and its time-reversed partner carries Hall conductance $\sigma_H = 8$ and central charge $c_- = 0$.\footnote{The $(1+1)$-D edge between the surface state and its time-reversed partner can be modeled as a chiral boson theory with $K$-matrix $K = \sigma^x$ and charge vector 
$\vec{t} = (2,2)$, which is equivalent to the edge theory of a bosonic integer quantum Hall state of charge
two bosons~\cite{LV2012,levin-senthil,VS2013}. The values $\sigma_H = 8$ and $c_- = 0$ follow immediately from this description.} Given this property, it follows from time-reversal symmetry that the $(1+1)$-D edge between this short-range entangled surface state and any $U(1)\rtimes\ZT$ symmetric topologically ordered surface state must carry Hall conductance $\sigma_H =4$ and central charge $c_- = 0$. Now consider the BTI$+e$ phase in a slab geometry with a symmetric topologically ordered surface state on the top surface of the slab and the short-range entangled surface state on the bottom surface of the slab. If we view this slab as a quasi-2D system, then its $(1+1)$-D edge (with the vacuum) again carries $\sigma_H = 4$ and $c_- = 0$. Hence by the formula (\ref{eta3fid}), the two dimensional topological order that describes the slab must have $\eta_{3,f} = -1$. It then follows that the topological order on the top surface must also have $\eta_{3,f} = -1$ since the rest of the slab, including the bottom surface, is short-range entangled. This establishes property $3$.

We now move on to discuss the $\mathbb{Z}_2$ anomaly related to the ETI phase. For this anomaly, we propose the indicator
\beq
	\nti = \frac{1}{2D}\sum_{a,b\in\C} d_a d_b S_{ab} e^{i2\pi q_a}e^{i2\pi q_b}\ , \label{eq:nti}
\eeq
where $S_{ab}$ denotes the $S$ matrix of the fermionic topological order $\mathcal{C}$ (as we review in Appendix~\ref{app:TO-with-e}, this $S$ matrix is actually degenerate, but this fact does not cause any problems in the definition of this indicator). In the remainder of this section, we will show that $\nti$ has all of the expected properties:
\begin{enumerate}
\item{$\nti$ can only take the values $+1$ and $-1$.}

\item{Any topological order that can be realized in strictly two dimensions has $\nti = +1$.}

\item{Any topological order that exists on the boundary of the ETI has $\nti = -1$.}
\end{enumerate}
In addition, we will show that all Abelian topological orders have $\nti=1$. This last property is an important consistency check on $\nti$ as there are general arguments that imply that any gapped, symmetry-preserving surface state for the ETI must possess \emph{non-Abelian} topological order~\cite{MKF2,CFV-2014,wang-potter-senthil-2,BNQ}.

\subsubsection{Examples}

In this section we compute the values of the indicators $\eta_{2,f}$, $\eta_{3,f}$, and $\nti$ for three important examples of fermionic topological orders, namely the ``ETMT$+e$", ``ECMC$+e$", and ``T-Pfaffian" topological orders. These examples are interesting because they are believed to appear on the surfaces of the three root phases in the $(\mathbb{Z}_2)^3$ classification of $(3+1)$-D fermionic SPTs with $U(1)\rtimes \ZT$ symmetry.

The ETMT$+e$ and ECMC$+e$ topological orders are Abelian topological orders which are obtained by appending the trivial electron $e$ to ETMT and ECMC topological orders built from charge $2$ bosons. In other words, the anyon content of these topological orders is $\{1,e\}\times\{1,E,M,\ep\}$, and so these topological orders have eight anyons in total including the electron. There is one important difference here compared to the bosonic case, which is that in the ECMC$+e$ state the $E$ and $M$ particles carry a charge of $1$ instead of $\frac{1}{2}$. This is because the ECMC topological order is built from charge $2$ bosons, so the $E$ and $M$ particles carry a charge of $\frac{1}{2} \cdot 2 = 1$.
For completeness, we also compute the values of our indicators for the ``EfMf$+e$" state, which is obtained from the bosonic EfMf state by appending the trivial electron. 

Moving on to the T-Pfaffian topological order~\cite{CFV-2014,BNQ}, this is the simplest of all of the non-Abelian topological orders that have been proposed as possible symmetry-preserving gapped surface states for the ETI. In particular, this topological order contains only 12 anyons including the electron. There are actually two versions of the T-Pfaffian state, known as the ``T-Pfaffian$_{\pm}$'' states, which differ slightly in their topological spins and $\T^2_a$ values. The T-Pfaffian$_{+}$ state is the version of the T-Pfaffian which is believed to appear on the surface of the non-interacting 
ETI~\cite{wang2015dual,metlitski-vishwanath,metlitski-s}. We review the anyon content and the detailed properties of the T-Pfaffian$_{\pm}$ topological orders in Appendix~\ref{app:TPf}.

In Table~\ref{tab:U1-rtimes-T-fermions} we record the values of the anomaly indicators $\eta_{2,f}$, $\eta_{3,f}$, and $\nti$ for the EfMf$+e$, ETMT$+e$, ECMC$+e$, and T-Pfaffian$_{\pm}$ topological orders. From the table we can see that $\eta_{2,f}$, $\eta_{3,f}$, and $\nti$ are independent, in the same sense that we discussed earlier in the bosonic case. We can also see that these three indicators can detect the anomalies in the ETMT$+e$, ECMC$+e$, and T-Pfaffian$_{\pm}$ topological orders, as well as in any combination of them. Assuming that the ETMT$+e$, ECMC$+e$, and T-Pfaffian$_{+}$ topological orders describe the surfaces of the three root states in the $(\mathbb{Z}_2)^3$ classification of $(3+1)$-D fermionic SPTs with $U(1)\rtimes \ZT$ symmetry, it then follows that $\eta_{2,f}$, $\eta_{3,f}$, and $\nti$ are a complete set of anomaly indicators for fermionic topological orders with $U(1)\rtimes\ZT$ symmetry.

Finally, we also mention here that the results in Table~\ref{tab:U1-rtimes-T-fermions} are consistent with several other results in the literature. In particular, we can see that T-Pfaffian$_{-}$ can be thought of as a weakly interacting mixture of the T-Pfaffian$_{+}$ and ETMT$+e$ states, as discussed in Ref.~\onlinecite{CFV-2014}. This can be seen from the table by noting that the value of any indicator for the T-Pfaffian$_{-}$ state is given by the product of the values of that indicator for the T-Pfaffian$_{+}$ and ETMT$+e$ states. A similar analysis shows that EfMf$+e$ can be thought of as a weakly interacting mixture of the ETMT$+e$ and ECMC$+e$ states (as discussed in Sec.~V.D of Ref.~\onlinecite{CFV-2014}). Our results are also consistent with the claim that the T-Pfaffian$_{+}$ state is not anomalous in the presence of $\ZT$ symmetry alone. 
However, we can see from the $\nti$ value that the T-Pfaffian$_{+}$ state \emph{is} anomalous in the presence of the full $U(1)\rtimes\ZT$ symmetry. This is consistent with the result that T-Pfaffian$_{+}$ is the version of the T-Pfaffian state which can appear on the surface of the non-interacting 
ETI~\cite{wang2015dual,metlitski-vishwanath,metlitski-s}. 

\begin{table}[t]
\begin{center}
\begin{tabular}{ c |c| c| c| c}
  & EfMf$+e$ & ETMT$+e$ & ECMC$+e$ & T-Pfaffian$_{\pm}$ \\ \hline
  $\eta_{2,f}$ & $-1$ & $-1$ & 1 & $\pm 1$ \\ \hline
  $\eta_{3,f}$ & $-1$ & 1 & $-1$ & 1 \\ \hline
  $\nti$ & 1 & 1 & 1 & $-1$ \\
\end{tabular} 
\end{center}
\caption{\label{tab:U1-rtimes-T-fermions}Values of the anomaly indicators $\eta_{2,f}$, $\eta_{3,f}$, and $\nti$ for the fermionic topological orders EfMf$+e$, ETMT$+e$, ECMC$+e$, and T-Pfaffian$_{\pm}$. Here, ETMT$+e$, ECMC$+e$, and T-Pfaffian$_{+}$ are three topological orders that can appear on the surfaces of the three root phases of the $(\mathbb{Z}_2)^3$ classification of fermionic SPTs with $U(1)\rtimes\ZT$ symmetry in $(3+1)$-D. We have also included the values of the anomaly indicators for the closely related EfMf$+e$, and T-Pfaffian$_{-}$ topological orders. In the right-most column the ``$\pm$" sign indicates that $\eta_{2,f}= 1$ for T-Pfaffian$_{+}$ and $\eta_{2,f}=-1$ for T-Pfaffian$_{-}$.}
\end{table}

\subsubsection{Physical picture and derivation of the indicator $\nti$}

We now present a derivation of the indicator $\nti$ and its properties. The derivation is very similar to the case of the indicator $\td{\eta}_3$ for the BTI, however, in the present case we have to deal with the subtleties of fermionic topological orders associated with the presence of the trivial electron $e$. As in the bosonic case, the derivation also uses three results about charge conserving and time-reversal symmetric fermionic topological orders. The first result, which we prove below, is that every $U(1)$ symmetric fermionic topological order $\C$ contains an Abelian anyon $f$ with the property that the mutual statistics between $f$ and any other anyon $a$ is given by
\begin{align}
e^{i\theta_{a,f}} = e^{i2\pi q_a}\ .
\end{align}
Unlike the bosonic case, however, in the fermionic case $f$ is only determined up to fusion with the electron, i.e., $f$ and $f\times e$ have the same mutual statistics $e^{i2\pi q_a}$ with all other anyons $a$. The second result, which we prove in Sec.~\ref{sec:bulk-bdy}, is that any $U(1)\rtimes\ZT$ symmetric fermionic topological order $\C$ that can appear on the surface of the ETI must have the property that $f$ has charge
\begin{align}
	q_f= \frac{1}{2} \text{ or }\frac{3}{2} \text{ (mod 2)}\ ,
\end{align}
where the ambiguity is again related to the fact that $f$ is only determined up to fusion with $e$. The third result, also proven in Sec.~\ref{sec:bulk-bdy}, is that any $U(1)\rtimes\ZT$ symmetric fermionic topological order $\C$ that can be realized in strictly two dimensions must have the property that $f$ has charge $q_f=0$ or $1$ (mod 2).  

These results suggest that a natural way to construct an anomaly indicator in the fermionic case is to define a function $\nti$ that takes as input a $U(1)\rtimes\ZT$ symmetric fermionic topological order $\C$, and returns as output the quantity $e^{i2\pi q_f}$. Note that with this definition it does not matter if we replace $f$ by $f\times e$ as we have $e^{i2\pi q_f}=e^{i2\pi q_{f\times e}}$. In addition, based on the properties we mentioned above, we are guaranteed that $\nti = -1$ for any topological order supported on the surface of the ETI, while $\nti = +1$ for any strictly two-dimensional topological order. We now discuss in detail how $\nti$ is constructed.

Our construction utilizes the quotient topological order $\hat{\C}$ that we define in Appendix~\ref{app:TO-with-e}, along with its associated $S$ matrix $\hat{S}$. Recall that the anyons in $\hat{\C}$ are equivalence classes $[a]$ of anyons in $\C$, where the equivalence relation is $a\sim b$ if $a= b\times e$, i.e., $a$ and $b$ correspond to the same anyon $[a]\in\hat{\C}$ if $a$ and $b$ differ by fusion with the electron. Since $\hat{\C}$ possesses a unitary $S$ matrix $\hat{S}$, we can apply to $\hat{\C}$ the same theorem from Ref.~\onlinecite{barkeshli1} that we used in the bosonic case. This theorem then implies that there is a unique anyon $[f]\in\hat{\C}$ whose mutual statistics with any other anyon $[a]\in\hat{\C}$ is equal to $e^{i2\pi q_{[a]}}$, where it is important to note that the charge of anyons $[a]\in\hat{\C}$ is determined modulo $1$ and not modulo $2$ (since $a$ and $a\times e$ are identified). Back in $\C$, this result implies that there are two anyons, $f$ and $f\times e$, whose mutual statistics with any other anyon $a\in\C$ is equal to $e^{i2\pi q_a}$, as we mentioned above. 

We now use the properties of $[f]$ and the \emph{unitary} $S$ matrix $\hat{S}$ of $\hat{\C}$ to construct a Kronecker delta function that can extract the anyon $[f]$ from a sum over $\hat{\C}$ (compare to the derivation of Eq.~\eqref{eq:NA-formula} in the bosonic case). The formula is
\beq
	\delta_{[b][f]}= \frac{1}{\hat{D}}\sum_{[a]\in\hat{\C}} d_{[a]} \hat{S}_{[a][b]} e^{i2\pi q_{[a]}}\ . \label{eq:NA-formula-fermions}
\eeq
We can then define an anomaly indicator function for the ETI as a sum over the quotient topological order $\hat{\C}$ as
\beq
	\nti = \frac{1}{\hat{D}}\sum_{[a],[b]\in\hat{\C}} d_{[a]} d_{[b]} \hat{S}_{[a][b]} e^{i2\pi q_{[a]}}e^{i2\pi q_{[b]}}\ ,
\eeq
where we inserted an extra factor of $d_{[b]}$, which is allowed since $d_{[f]}=1$. By construction, this sum evaluates to 
\begin{align*}
	\nti= e^{i2\pi q_{[f]}}= e^{i2\pi q_f}\ ,
\end{align*}
where the second equality follows from the fact that $f$ and $f\times e$ differ by a charge of $1$. Finally, now that we possess this formula for $\nti$, we can use the symmetries of the original $S$ matrix $S_{ab}$ of the theory $\C$ (which are recorded in Eq.~\eqref{eq:S-symmetries}) to write $\nti$ as a sum over the entire 
topological order $\C$, instead of as a sum over the quotient $\hat{\C}$. We find that
\begin{align*}
	\nti = \frac{1}{2D}\sum_{a,b\in\C} d_a d_b S_{ab} e^{i2\pi q_a}e^{i2\pi q_b}\ ,
\end{align*}
which is exactly the original form of this indicator that we presented in Eq.~\eqref{eq:nti}.

\subsubsection{Proof that the indicator $\nti$ takes the values $\pm 1$}

We now prove that $\nti$ can only take on the values $\pm 1$ for fermionic topological orders with $U(1)\rtimes\ZT$ symmetry. To prove this we show that in the presence of time-reversal symmetry, the charge of the flux anyon $f$ can only take the values $q_f= 0, \frac{1}{2}, 1 \text{ or } \frac{3}{2} \pmod{2}$. Since the anomaly indicator $\nti$ is equal to $e^{i2\pi q_f}$, it immediately follows from this that $\nti=\pm 1$ only.  

To prove the above property of $q_f$ we study the mutual statistics $\theta_{f,a}$ of $f$ with another anyon $a$ (this is always a phase since $f$ is Abelian). We have $\theta_{f,a}= 2\pi q_a$ (mod $2\pi$) by the properties of the flux anyon $f$. In addition, the transformation properties of statistical phases under time-reversal imply that $\theta_{\T(f),a}= -2\pi q_a$. By adding the equations for $f$ and $\T(f)$ we find that $\theta_{f,a}+\theta_{\T(f),a}=\theta_{f\times \T(f),a}= 0$ (mod $2\pi$), which implies that $f\times \T(f)$ has trivial mutual statistics with all anyons $a$. Then it must be the case that either (i) $\T(f)= \ov{f}$, or (ii) $\T(f)= \ov{f}\times e$, since by assumption only $1$ and $e$ have trivial mutual statistics with all other anyons. Now the charges of anyons obey the two relations
\begin{align}
	q_a = -q_{\ov{a}} \ , \quad \quad q_{a} = q_{\T(a)}\ ,
	\label{qatarel}
\end{align}
where both equations hold modulo $2$. Then in case (i) these equations together imply that $2 q_f=0 \pmod{2}$ so that $q_f=0$ or $1 \pmod{2}$, while in case (ii) these equations imply that $2q_f= 1 \pmod{2}$ so that $q_f= \frac{1}{2}$ or $\frac{3}{2} \pmod{2}$. 

\subsubsection{Proof that $\nti=1$ for Abelian topological orders}

We now prove that $\nti=1$ for any Abelian fermionic topological order
$\C$. This fact is important because it demonstrates that our proposed indicator $\nti$ is consistent with 
general arguments~\cite{MKF2,CFV-2014,wang-potter-senthil-2,BNQ} implying that the boundary of the ETI does not support a symmetric topological order in which all quasiparticles have Abelian statistics. 

Our proof begins with a result from the previous section, namely the result that the charge $q_f$ of the flux anyon $f$ obeys either (i) $2q_f= 0 \pmod{2}$, or (ii) $2 q_f = 1 \pmod{2}$. Next, we show that in all Abelian theories with a trivial electron $e$ it must actually be the case that $2q_f=0 \pmod{2}$, which then implies $\nti = 1$. To show this, we first recall that any Abelian fermionic topological order $\C$ admits a decomposition
\begin{align*}
	\C = \{a_1,\dots, a_n,a_1\times e,\dots, a_n\times e\}\ ,
\end{align*}
where the subset 
\begin{align*}
	\C' = \{a_1,\dots, a_n\}
\end{align*}
has the very important property that it is closed under fusion (here $2n$ is equal to the total number of anyons in 
$\C$). [For a proof, see Corollary A.19 of 
Ref.~\onlinecite{arXiv:0906.0620}]. In this decomposition, $\C'$ will contain the identity anyon $1$, and so the electron $e$ is then not in $\C'$.  
One more important point is that although $\C'$ is closed under fusion, it may not be closed under the action of time-reversal. 

Within the set $\C'$ there exists a new function $\phi':\C'\to \mathbb{R}$, defined by $\phi'_a = -\pi q_a$, which has the property that
\beq
	e^{i\phi'_a}e^{i\phi'_b}= e^{i\phi'_c}\ 
\eeq
for any $a,b,c\in\C'$ with $a\times b= c$. This is true because the charges of all anyons are defined modulo $2$ in theories with a trivial electron $e$. Then by the same theorem from Ref.~\onlinecite{barkeshli1} that we applied in the bosonic case, there exists an Abelian anyon $g\in\C'$ such that
\beq
	e^{i\theta_{g,a}}= e^{i \pi q_a}
\eeq
for all $a\in\C'$. From this equation, it follows that either $g\times g=f$, or $g\times g= f\times e$. 

To complete the proof, consider the anyon $g \times \T(g)$. Notice that
\begin{align}
[g \times \T(g)] \times [g \times \T(g)] &= f \times \T(f) \nonumber \\
&= 1 \text{ or } e
\label{fusionprodgtg}
\end{align}
where the second equality follows from the discussion above Eq.~(\ref{qatarel}). In fact, it is not possible for the above fusion product (\ref{fusionprodgtg}) to be $e$. To see this, observe that $g \times \T(g)$ is either an element of $\C'$ or $\C' \times e$. If $g \times \T(g) \in \C'$, then the fusion product (\ref{fusionprodgtg}) cannot be $e$ since $\C'$ is closed under fusion and $e$ does not belong to $\C'$. Likewise, if $g \times \T(g) \in \C' \times e$, then the fusion product again cannot be $e$ by the same reasoning. 

We are now finished: the only consistent possibility is $f \times \T(f) = 1$, which means that $2q_f= 0 \pmod{2}$, as we wished to show.

This proof raises a question: why is there no anyon $g$ in $\C$ such that $\theta_{g,a}= \pi q_a$ for all $a\in\C$, and not only for all $a\in\C'$? After all, the charges $q_a$ for the anyons in $\C$ obey 
$q_a+q_b= q_c$ mod $2$ for any $a,b,c\in\C$ such that $N^c_{ab}\neq 0$, and this implies that 
$e^{i\phi'_a}e^{i\phi'_b}= e^{i\phi'_c}$ holds for all $a,b,c\in\C$, not just in $\C'$. The answer to this can be seen by looking at the case where $a$ is the electron $e$ (which is not in $\C'$), which would give $\theta_{g,e}= \pi$, i.e., if $g$ was defined for all of $\C$ then there would be an anyon $g\in\C$ with nontrivial mutual statistics with the electron. But this is a contradiction since one of our original assumptions about $\C$ was that the electron has trivial mutual statistics with all other anyons, as expressed in Eq.~\eqref{eq:locality}.

\section{Universal properties of topological insulator surfaces}
\label{sec:bulk-bdy}

\subsection{Bosonic topological insulator}
\label{sec:btideriv}
\begin{figure}[t]
  \centering
    \includegraphics[width= .5\textwidth]{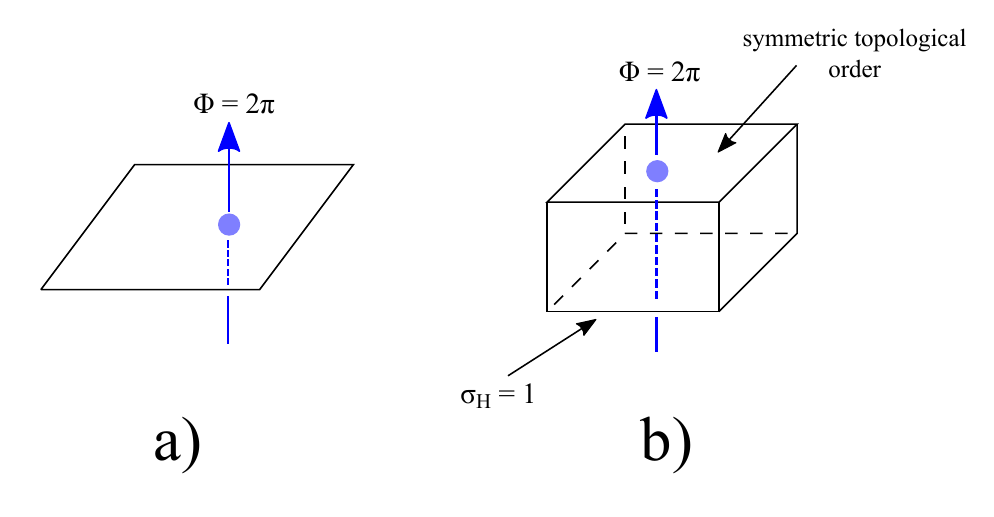} 
\caption{a) Threading a thin $2\pi$ flux tube in a two-dimensional system. b) Threading a thin $2\pi$ flux tube through a
three-dimensional bosonic topological insulator slab featuring a symmetric topologically ordered top surface and a 
symmetry-breaking (and short-range entangled) bottom surface with $\sigma_H=1$. In both figures the blue circle represents
the location of the flux anyon $f$.}
\label{fig:flux-BTI}
\end{figure}

In this section, we prove two claims about bosonic topological insulator surface states which we asserted in Sec.~\ref{sec:U1-rtimes-T}. Before presenting the proofs, we first review an important fact that comes into both arguments: for any charge conserving, two-dimensional gapped bosonic lattice model, the exchange statistics of the $2\pi$ flux anyon $f$ is related to the Hall conductivity $\sigma_H$ via
\begin{align}
e^{i\theta_f} = e^{i \pi\sigma_{H}}\ .
\label{statcond}
\end{align}
To derive Eq.~\eqref{statcond}, imagine starting in the ground state of the lattice model and then adiabatically inserting $2\pi$ magnetic flux through some finite region of the 2D plane (Fig.~\ref{fig:flux-BTI}a). It is easy to see that this process nucleates the flux anyon $f$ in the region where the magnetic field is applied: indeed, if we braid any other anyon $a$ around this region, the resulting Berry phase is the Aharonov-Bohm phase $e^{i2\pi q_a}$, which is precisely the definition of the flux anyon.
 
Now imagine that we create \emph{two} flux anyons using these flux insertion processes. It follows from the usual Hall response arguments that each flux insertion process pumps charge $\sigma_{H}$ into the region where the magnetic field is applied. Therefore, the Berry phase associated with braiding one flux anyon around the other is $e^{i2\pi \sigma_H}$. Similar reasoning implies that the Berry phase associated with \emph{exchanging} two flux anyons is $e^{i\pi \sigma_H}$. This completes the derivation of (\ref{statcond}).

With Eq.~\eqref{statcond} in hand, we are now ready to prove our first claim: any topological order $\mathcal{C}$ that can be realized by a strictly two-dimensional, $U(1)$ conserving, time-reversal symmetric bosonic lattice model must have the property that the flux anyon $f$ is a boson. To see this, note that any lattice realization must have a vanishing Hall conductivity, $\sigma_H = 0$, due to time-reversal symmetry. Applying (\ref{statcond}), we conclude that if $\mathcal{C}$ has a two-dimensional lattice realization, then the flux anyon must be a boson.

We now move on to the second claim: any symmetric topological order $\mathcal{C}$ that can be realized on the surface of the BTI must have the property that the flux anyon $f$ is a \emph{fermion}. Our proof relies on the following fundamental property of the BTI: its surface can host a gapped, $U(1)$ conserving, time-reversal breaking, short-range entangled state with a surface Hall conductivity $\sigma_H = 1$.

The first step of the proof is to consider a BTI in a slab geometry in which the bottom surface hosts the short-range entangled $\sigma_H = 1$ state, while the top surface hosts the symmetric topological order $\mathcal{C}$ (Fig.~\ref{fig:flux-BTI}b). We will think of this slab as a (quasi) two-dimensional bosonic lattice model.

The second step is to observe that the slab has a Hall conductivity of $\sigma_H = 1$ since it is time-reversal symmetric apart from the bottom surface. Therefore, according to (\ref{statcond}), the $2\pi$ flux anyon $f$ for the slab has exchange statistics $e^{i\theta_f} = -1$, i.e. $f$ is a fermion.

Finally, we observe that the $2\pi$ flux anyon for the slab belongs to the same superselection sector as the $2\pi$ flux anyon associated with the topological order $\mathcal{C}$ on the top surface --- that is, the two flux anyons can be transformed into one another by a unitary operator that is local in the 2D sense. To see this, note that the both the bulk and the lower surface of the slab are short-range entangled and therefore the only anyon excitations of the slab are those that live on the top surface. It follows that flux anyon for the top surface and the flux anyon for the slab have the same mutual statistics with every other anyon excitation in the slab, namely $e^{i\theta_{f,a}} = e^{i2\pi q_a}$. Hence, according to the braiding non-degeneracy property~\cite{kitaevhoneycomb} of bosonic topological orders, it must be possible to transform the two flux anyons into one another using a unitary operator that is local in the 2D sense.

We are now done: it follows from the above observation that the flux anyon for the top surface has the same exchange statistics as the flux anyon for the slab, i.e. it is a fermion. This is what we wanted to show.

\subsection{Electron topological insulator}

We now move on to prove two claims about electron topological insulator surface states which we asserted in Sec.~\ref{sec:U1-rtimes-T}. The proofs are similar to those in the bosonic case discussed above.

To begin, we review a result that plays an important role in our derivations: for any charge conserving, two-dimensional gapped fermionic lattice model, the charge of the $2\pi$ flux anyon $f$ is related to the Hall conductivity $\sigma_H$ via
\begin{align}
q_f = \sigma_{H} \text{ or } 1 +\sigma_H  \pmod{2}
\label{chgcond}
\end{align}
To derive Eq.~\eqref{chgcond}, observe that, just like the bosonic case, we can construct the $2\pi$ flux anyon by adiabatically inserting $2\pi$ magnetic flux through some finite region of the 2D plane. This process pumps charge $\sigma_H$ into the region where the magnetic field is applied. This implies (\ref{chgcond}) since the $2\pi$ flux anyon is only determined up to fusion with the electron, which carries charge $1$.

With Eq.~\eqref{chgcond} in hand, we are now ready to prove our first claim: any topological order $\mathcal{C}$ that can be realized by a strictly two-dimensional, $U(1)$ conserving, time-reversal symmetric fermionic lattice model must have the property that the flux anyon $f$ has charge $q_f = 0$ or $1 \pmod{2}$. To see this, note that any lattice realization has a vanishing Hall conductivity, $\sigma_H = 0$, due to time-reversal symmetry. Applying (\ref{statcond}) we conclude that if $\mathcal{C}$ has a lattice realization, then the flux anyon must carry charge $0$ or $1 \pmod{2}$. 

We now move on to the second claim: any symmetric topological order $\mathcal{C}$ that can be realized on the surface of the ETI must have the property that the flux anyon $f$ has charge $q_f = 1/2$ or $3/2 \pmod{2}$. Our proof relies on the following universal property of the electron topological insulator: its surface can host a gapped, $U(1)$ conserving, 
time-reversal breaking, short-range entangled state with a surface Hall conductivity $\sigma_H = 1/2$.

The first step of the proof is to consider an ETI in a slab geometry in which the bottom surface hosts the short-range entangled $\sigma_H = 1/2$ state, while the top surface hosts the symmetric topological order $\mathcal{C}$. We will think of this slab as a (quasi) two-dimensional electronic lattice model.

Next, notice that the slab has a Hall conductivity of $\sigma_H = 1/2$, since it is time-reversal symmetric apart from the bottom surface. Therefore, according to (\ref{statcond}), the $2\pi$ flux anyon $f$ for the slab has charge $q_f = 1/2$ or $3/2 \pmod{2}$.

Finally, observe that the $2\pi$ flux anyon for the slab belongs to the same superselection sector as the $2\pi$ flux anyon associated with the topological order $\mathcal{C}$ on the top surface --- that is, the two flux anyons can be transformed into one another (up to fusion with an electron) by a unitary operator that is local in the 2D sense. This follows by the same arguments as in the bosonic case. 

This completes the proof: it follows from the above observation that the flux anyon for the top surface has the same charge as the flux anyon for the slab, i.e $q_f = 1/2$ or $3/2 \pmod{2}$.

\subsection{Alternative derivation}

\begin{figure}[t]
  \centering
    \includegraphics[width= .5\textwidth]{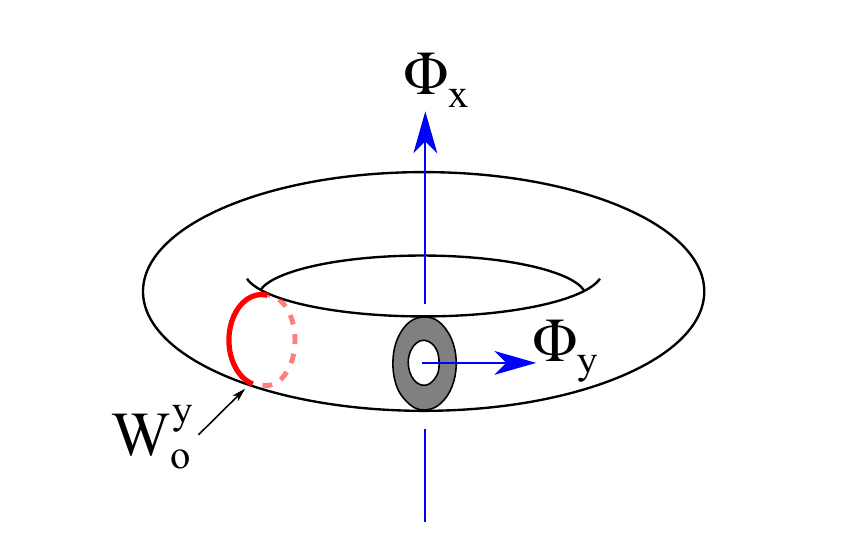} 
\caption{A 3D electronic topological insulator in a thickened torus (or Corbino donut) geometry with fluxes $\Phi_x$ and $\Phi_y$ threading the $x$- and $y$-cycles of the torus. Also pictured is the string operator $W^y_o$ (the red line) wrapping the $y$-cycle of the outer surface of the torus.}
\label{fig:corbino}
\end{figure}

Before closing this section, we now give an alternative derivation of the fact that any symmetric topological order $\mathcal{C}$ that can be realized on the surface of the ETI must have the property that the flux anyon has charge $q_f = 1/2$ or $3/2 \pmod{2}$. This derivation relies on another universal property of electronic topological insulators which was alluded to in Ref.~\onlinecite{fu-kane2007} and explained in more detail in Ref.~\onlinecite{Levin-FTI}. To explain this property, consider an ETI in a thickened torus (also known as a `Corbino donut') geometry in which both surfaces of the thickened torus are $U(1)$ and time-reversal symmetric. This thickened torus geometry is shown in Fig.~\ref{fig:corbino}. Suppose that:
\begin{enumerate}
\item{The ground state (or one of the degenerate ground states) contains an even number of electrons.}
\item{The ground state does not have a local Kramers degeneracy (i.e. Kramers doublets) on either surface.}
\end{enumerate}
Starting in this ground state, which we will call $|\Psi_{0,0}\rangle$, we can consider three processes in which either $0$ or $\pi$ flux is inserted adiabatically through the two holes of the torus. There are three states that we can construct by such an adiabatic flux insertion, which we can label as $|\Psi_{\pi,0}\rangle$, $|\Psi_{0,\pi}\rangle$, and $|\Psi_{\pi,\pi}\rangle$. The universal property of an ETI is that an \emph{odd} number of these states carry a local Kramers degeneracy on the two surfaces of the thickened torus.

The first step of the proof is to consider the thickened torus geometry of the kind described above, where both surfaces of the thickened torus host the symmetric topological order $\mathcal{C}$. Consider the
state $|\Psi_{\pi,0}\rangle$ obtained by adiabatically inserting $\pi$ flux through the `$x$-cycle' of the torus. According to the definition given in Ref.~\onlinecite{levin-stern}, we can compute the local Kramers degeneracy $\mathcal{T}^2_{\pi,0} = \pm 1$ carried by $|\Psi_{\pi,0}\rangle$ if we can find operators $S_o, S_i$ that are supported on the outer and inner surfaces of the thickened torus with the property that
\begin{align*}
S_o S_i |\Psi_{\pi,0}\rangle \propto \mathcal{T} |\Psi_{\pi,0}\rangle
\end{align*}
and with $\|S_o |\Psi_{\pi,0}\rangle \| = \|S_i |\Psi_{\pi,0}\rangle \| = 1$.  Once we do this, we can read off the Kramers degeneracy $\T^2_{\pi,0}$ as follows: 
\begin{align}
\mathcal{T} S_o \mathcal{T} S_o |\Psi_{\pi,0}\rangle =  \mathcal{T}^2_{\pi,0} |\Psi_{\pi,0}\rangle
\label{kramersdef}
\end{align}
(The same equation also holds if we replace $S_o \rightarrow S_i$).

To construct $S_o, S_i$, we make three observations. First, $\mathcal{T} |\Psi_{\pi,0}\rangle = |\Psi_{-\pi,0}\rangle$ where $|\Psi_{-\pi,0}\rangle$ is the state obtained by starting in $|\Psi_{0,0}\rangle$ and adiabatically inserting $-\pi$ flux through the `$x$-cycle' of the torus. Second, we observe that $|\Psi_{-\pi,0}\rangle$ can be obtained by starting in $|\Psi_{\pi,0}\rangle$ and adiabatically inserting $-2\pi$ flux through the `$x$-cycle' of the torus. Finally, we note that this flux insertion process has the same effect as braiding the $2\pi$ flux anyon $f$ around the $y$-cycle, on both surfaces of the thickened torus. Putting this this all together, we conclude that $S_o, S_i$ can be chosen to be these flux anyon braiding processes on the two surfaces: that is,
\begin{align*}
S_o = W^y_o, \quad \quad S_i = W^y_i 
\end{align*}
where $W^y_o$ and $W^y_i$ denote the string-like operators that braid the flux anyon $f$ around the $y$-cycle of the outer and inner surfaces of the thickened torus. (The operator $W^y_o$ is represented by the thick red line in Fig.~\ref{fig:corbino}).

Substituting $S_o = W^y_o$ into (\ref{kramersdef}), we deduce that the local Kramers degeneracy of $|\Psi_{\pi,0}\rangle$ can be obtained from
\begin{align}
\mathcal{T} W^y_o \mathcal{T} W^y_o |\Psi_{\pi,0}\rangle =  \mathcal{T}^2_{\pi,0} |\Psi_{\pi,0}\rangle
\label{tpi0}
\end{align}
Similarly, the local Kramers degeneracy of  $|\Psi_{0,\pi}\rangle$ can be obtained from
\begin{align}
\mathcal{T} W^x_o \mathcal{T} W^x_o |\Psi_{0,\pi}\rangle =  \mathcal{T}^2_{0,\pi} |\Psi_{0,\pi}\rangle
\label{t0pi}
\end{align}
where $W^x_o$ denotes the string-like operator that braids the flux anyon $f$ around the \emph{$x$-cycle} of the outer surface of the torus. By the same reasoning, the local Kramers degeneracy of $|\Psi_{\pi,\pi}\rangle$ is given by
\begin{align}
\mathcal{T} W^x_o W^y_o \mathcal{T} W^x_o W^y_o |\Psi_{\pi,\pi}\rangle =  \mathcal{T}^2_{\pi,\pi} |\Psi_{\pi,\pi}\rangle
\label{tpipi}
\end{align}
To proceed further, note that 
\begin{align*}
\mathcal{T} W^y_o \mathcal{T} \propto \bar{W}^y_o
\end{align*}
where $\bar{W}^y_o$ denotes the string operator that braids $\mathcal{T}(f)$ around the $y$-cycle of the outer surface of the thickened torus. Also recall that $\mathcal{T}(f) = \bar{f}$ or $\bar{f} \times e$. It follows that $W^x_o, \bar{W}^y_o$ commute with one another up to a phase, which is precisely the mutual statistics of the $f$ anyon with its antiparticle $\bar{f}$:
\begin{align*} 
 W^x_o \bar{W}^y_o = e^{-i2\pi q_f} \bar{W}^y_o W^x_o
\end{align*}

Putting this all together, we derive
\begin{align}
&\mathcal{T} W^x_o W^y_o \mathcal{T} W^x_o W^y_o |\Psi_{\pi,\pi}\rangle \nonumber \\
&= (\mathcal{T} W^x_o \mathcal{T}) (\mathcal{T} W^y_o \mathcal{T}) W^x_o W^y_o |\Psi_{\pi,\pi}\rangle \nonumber \\
&= e^{i2\pi q_f} (\mathcal{T} W^x_o \mathcal{T}) W^x_o (\mathcal{T} W^y_o \mathcal{T}) W^y_o |\Psi_{\pi,\pi}\rangle \nonumber \\
&= e^{i2\pi q_f} \mathcal{T}^2_{0,\pi}  \mathcal{T}^2_{\pi,0} |\Psi_{\pi,\pi}\rangle
\label{tpipi2}
\end{align}
Here, in the last line we used the following two identities:
\begin{align*}
\mathcal{T} W^y_o \mathcal{T} W^y_o |\Psi_{\pi,\pi}\rangle =
\mathcal{T}^2_{\pi,0} |\Psi_{\pi,\pi}\rangle
\end{align*}
and
\begin{align*}
\mathcal{T} W^x_o \mathcal{T} W^x_o |\Psi_{\pi,\pi}\rangle =
\mathcal{T}^2_{0,\pi} |\Psi_{\pi,\pi}\rangle
\end{align*}
To derive the first identity, notice that the operator $\mathcal{T} W^y_o \mathcal{T} W^y_o$ describes a braiding process that does not move any (net) charge around the $y$th cycle of the torus. This means that $\mathcal{T} W^y_o \mathcal{T} W^y_o$ must multiply $|\Psi_{\pi,\pi}\rangle$ by the same phase factor as $|\Psi_{\pi,0}\rangle$: in general, the effect of \emph{any} $y$-cycle braiding process on $|\Psi_{\pi,\pi}\rangle$ is the same as on $|\Psi_{\pi,0}\rangle$ except for an Aharonov-Bohm phase factor $e^{i\pi Q}$ where $Q$ is the total charge that is braided. Combining this observation with Eq. (\ref{tpi0}) gives the first identity. The second identity follows from Eq. (\ref{t0pi}) by the same reasoning.
  
Comparing (\ref{tpipi2}) with (\ref{tpipi}), we conclude that
\begin{align*}
\mathcal{T}^2_{\pi,\pi} = e^{i2\pi q_f} \mathcal{T}^2_{0,\pi}  \mathcal{T}^2_{\pi,0}
\end{align*}
Given that an odd number of the $\mathcal{T}^2$'s have to be $-1$ for an ETI, we deduce that $e^{i2\pi q_f} = -1$. Thus the charge of the flux anyon must be $q_f = 1/2$ or $3/2 \pmod{2}$. This is what we wanted to show.

\section{Anomaly indicators for systems with $U(1)\times\mathbb{Z}_2^T$ symmetry}
\label{sec:U1-times-T}

For most of this paper we have been concerned with bosonic or fermionic topological orders with $U(1)\rtimes \ZT$ symmetry. In this final section we switch gears and focus on the case of $U(1)\times \ZT$ symmetry. In this case, the $U(1)$ part of the symmetry group is usually interpreted as conservation of one component of a spin degree of freedom in a spinful system --- for example conservation of the $z$-component of spin. This interpretation follows from the fact that the Cartesian product structure of the full symmetry group, combined with the fact that time-reversal is anti-unitary, implies that time-reversal \emph{negates} the $U(1)$ charge. It follows that in topological orders with this symmetry group, the charges $q_a$ of the anyons must behave under time-reversal as
\beq
	q_{\T(a)}= -q_a\ , \label{eq:q-U1-times-T}
\eeq
where this equation should be interpreted modulo $1$ in the bosonic case and modulo $2$ in the fermionic case.

\subsection{Bosonic case}

\subsubsection{An additional indicator $\eta_4$}

Bosonic SPT phases with $U(1)\times\ZT$ symmetry in $(3+1)$-D are believed to have a $(\mathbb{Z}_2)^4$ classification~\cite{VS2013,kapustin1} (the group cohomology approach~\cite{CGLW2013} gives only 
$(\mathbb{Z}_2)^3$). Of these four $\mathbb{Z}_2$ factors, the first three have essentially the same physical characterization as the phases that appear in the $U(1)\rtimes\ZT$ case.\footnote{One way to understand this result is to note that the effective actions used to classify the phases corresponding to the first three $\mathbb{Z}_2$ factors are the same for both symmetry groups~\cite{kapustin1}.} In particular, the first two $\mathbb{Z}_2$ factors correspond to phases that require only $\ZT$ symmetry, and the third $\mathbb{Z}_2$ factor is again characterized by the property that the flux anyon on the surface is a fermion. It follows that the topological orders that can appear on the surfaces of the root phases for the first three $\mathbb{Z}_2$ factors can be detected by the same three indicators $\eta_1$, $\eta_2$, and $\eta_3$ that we already discussed in the $U(1)\rtimes\ZT$ case. One nontrivial point here is that our previous proof that $\eta_3 = \pm 1$ was only valid for the $U(1)\rtimes \ZT$ case. Below we prove that $\eta_3=\pm 1$ in the $U(1)\times\ZT$ case as well, where we have $q_{\T(a)}=-q_a$ instead of $q_{\T(a)}=q_a$.  

It remains to find an indicator that can detect the anomaly associated with the final $\mathbb{Z}_2$ factor in the $(\mathbb{Z}_2)^4$ classification. For this purpose we propose the indicator
\beq
	\eta_4= \frac{1}{D}\sum_{a,b\in\C}d_a S_{ab} e^{i2\pi q_a} \T^2_b\ ,
\eeq
where all symbols have the same definition as in the case of bosonic topological orders with $U(1)\rtimes\ZT$ symmetry that we covered in Sec.~\ref{sec:U1-rtimes-T-bosons}. In this section we prove that $\eta_4$ has the following two properties:
\begin{enumerate}
\item{$\eta_4$ can only take the values $+1$ and $-1$.}
\item{Any topological order that can be realized in strictly two dimensions has $\eta_4 = +1$.}
\end{enumerate}
In fact, property $1$ will be built into the construction of $\eta_4$, and so it will not require a proof. Property $2$, however, does require a proof, and we supply this proof in Sec.~\ref{sec:eta4deriv}. We also conjecture that $\eta_4$ has the third property:
\begin{enumerate}
\setcounter{enumi}{2}
\item Any topological order that exists on the boundary of the fourth root state in the $(\mathbb{Z}_2)^4$ classification has $\eta_4 = -1$.
\end{enumerate}
This conjecture is based on the fact that the indicator $\eta_4$ is designed to capture a certain physical characterization, put forward in Refs.~\onlinecite{VS2013,senthil-review}, of the nontrivial SPT phase corresponding to the final $\mathbb{Z}_2$ factor in the $(\mathbb{Z}_2)^4$ classification of bosonic SPT phases with $U(1)\times\ZT$ symmetry in $(3+1)$-D. If this physical characterization is indeed correct, then property $3$ will be true. We explain this characterization in detail below when we discuss the physical picture and derivation of $\eta_4$. 

\subsubsection{Examples}

We now compute the indicators $\{\eta_1,\eta_2,\eta_3,\eta_4\}$ for the EfMf, ETMT, and ECMC topological orders (discussed previously) as well as an additional state known as the ``ETMC" state~\cite{kapustin1}, which is defined as follows. The anyon content of the ETMC state is the same as in the usual toric code model, but in this state the $E$ particle is a Kramers doublet, the $M$ particle is a Kramers singlet and carries charge $\frac{1}{2}$, and the $\ep$ particle is a Kramers doublet with charge $\frac{1}{2}$. This state has been argued to occur on the surface of the fourth root phase in the $(\mathbb{Z}_2)^4$ classification of bosonic SPT phases with $U(1)\times\ZT$ symmetry in $(3+1)$-D. Likewise, the EfMf, ETMT, ECMC states are believed to describe the surfaces of the first three root phases in this classification.

In Table~\ref{tab:U1-times-T-bosons} we display the values of $\{\eta_1,\eta_2, \eta_3, \eta_4\}$ for the four bosonic topological orders EfMf, ETMT, ECMC, and ETMC. We can draw several conclusions from these results. First, we can see that these four indicators are independent, in the sense that the values displayed in the table rule out the possibility that one of the four indicators can be written as a product of two or more of the others. The table also shows that these indicators can successfully detect the anomalies in all four of these topological orders, and in any mixture of them. Assuming that these four topological orders appear on the surfaces of the four root SPT phases, it follows that $\{\eta_1,\eta_2,\eta_3,\eta_4\}$ provide a complete set of anomaly indicators for bosonic topological orders with $U(1)\times\ZT$ symmetry.

Finally, we also mention here that a slightly different version of the ETMC state was considered in Table V of Appendix D in Ref.~\onlinecite{VS2013}. In this alternative version of ETMC, which we may call ETMC$'$, the $E$ particle is a charge $1/2$ Kramers doublet and the $M$ particle is a neutral Kramers doublet. As a result, $\ep$ is a charge $1/2$ Kramers singlet. We have checked that this ETMC$'$ state has the same anomaly $\eta_4=-1$ as the ETMC state from Ref.~\onlinecite{kapustin1} that we discussed above. In addition, we find that $\eta_1=1$, $\eta_2=-1$, and $\eta_3=1$ for this state. It follows from these values that ETMC$'$ has the same anomalies as a weakly interacting mixture of the ETMC and ETMT states, and so the ETMC$'$ state is not independent of the other states that we have already discussed.

\begin{table}[t]
\begin{center}
\begin{tabular}{ c |c| c| c | c}
  & EfMf & ETMT & ECMC & ETMC \\ \hline
$\eta_1$ & $-1$ & 1 & 1 & 1\\ \hline
$\eta_2$ & $-1$ & $-1$ & 1 & 1\\ \hline
$\eta_3$ & $-1$ & 1 & $-1$ & 1\\ \hline
$\eta_4$ & 1 & 1 & 1 & $-1$ \\
\end{tabular} 
\caption{\label{tab:U1-times-T-bosons} Values of the indicators $\eta_1$, $\eta_2$, $\eta_3$, and $\eta_4$ for the EfMf, ETMT, ECMC, and ETMC topological orders which can appear on the surfaces of the four root phases in the $(\mathbb{Z}_2)^4$ classification of bosonic SPTs with $U(1)\times\ZT$ symmetry in $(3+1)$-D.}
\end{center}
\end{table}

\subsubsection{Proof that $\eta_3=\pm 1$ in the case of $U(1)\times \ZT$ symmetry}

We now prove that the indicator $\eta_3$ can take only the values $\pm 1$ in the case of $U(1)\times \ZT$ symmetry. We start by proving that $\eta_3$ is real. To do this we first compute the complex conjugate of $\eta_3$, which is $\eta^*_3= \frac{1}{D}\sum_{a\in\C} d_a^2  e^{-i\theta_a}  e^{-i2\pi q_a}$. Next, we make a change of summation variable by writing $a=\T(b)$ for some anyon $b$. Since $\theta_{\T(b)}=-\theta_b$ and $q_{\T(b)}= -q_b$ in the $U(1)\times \ZT$ case, we immediately find that $\eta_3$ is real in this case. The proof that $\eta_3=\pm 1$ then uses this fact combined with the relation $\td{\eta}_3=\eta_1\eta_3$ (which continues to hold in the $U(1)\times \ZT$ case), and is identical to the proof presented at the end of Sec.~\ref{sec:boson-physical-picture}. Note that when proving that $\td{\eta}_3=\eta_1\eta_3$ in the $U(1)\times \ZT$ case, slightly different manipulations are needed to get from Eq.~\eqref{eq:eta3-eqn} to the final result. In particular, two changes of the summation variable are needed. First, we set $b=\ov{c}$ in that equation and use $\theta_{\ov{c}}=\theta_c$ and $q_{\ov{c}}=-q_c$. Next, we set $c=\T(a)$ and use $\theta_{\T(a)}=-\theta_a$ and $q_{\T(a)}=-q_a$ to arrive at the final result.

\subsubsection{Physical picture and derivation of the indicator $\eta_4$}
\label{sec:eta4deriv}

We now present the derivation of the indicator $\eta_4$ and its properties. The derivation is based on a physical insight about the properties of the flux anyon $f$ in a system with $U(1)\times \ZT$ symmetry. The key insight is that in a system with $U(1)\times\ZT$ symmetry, the structure of the symmetry group actually \emph{requires} the flux anyon $f$ to be time-reversal invariant (we provide a proof of this below). It is then possible for $f$ to be a Kramers singlet or a Kramers doublet, with the Kramers doublet case corresponding to the anomaly~\cite{VS2013,senthil-review}. Since this property of $f$ \emph{does not} hold in the $U(1)\rtimes\ZT$ case, this anomaly must correspond to the extra $\mathbb{Z}_2$ factor (compared to the $U(1)\rtimes\ZT$ case) in the $(\mathbb{Z}_2)^4$ classification of $(3+1)$-D bosonic SPT phases with $U(1)\times\ZT$ symmetry.

This physical idea immediately suggests a way to construct an anomaly indicator to detect the anomaly associated with the remaining $\mathbb{Z}_2$ factor: we define a function $\eta_4$ that takes as input a bosonic topological order with $U(1)\times\ZT$ symmetry and returns the value $\T^2_f$ for the flux anyon $f$. We are then guaranteed that $\eta_4=-1$ on the surface of the nontrivial SPT phase with $U(1)\times\ZT$ corresponding to the fourth $\mathbb{Z}_2$ factor in the classification. Also, \emph{by construction}, $\eta_4$ can only take on the two values $\pm 1$, so we will not need a separate proof that $\eta_4$ can take only these two values. We can also see that $\eta_4=1$ for any $U(1)\times \ZT$ symmetric bosonic topological order that can be realized in strictly two dimensions: to prove this, recall that the flux anyon $f$ can be created by a process in which $2\pi$ flux is adiabatically inserted in some finite region of the 2D plane (see Sec.~\ref{sec:btideriv}). This flux insertion process is manifestly invariant under $\ZT$ (since the flux for the $U(1)$ gauge field is $\ZT$ invariant), so the final state $f$ is also $\ZT$ invariant. We conclude that there exists a microscopic state that belongs to the same superselection sector as $f$ and that is $\ZT$ invariant. It follows that $f$ must be a Kramers singlet, i.e. $\eta_4 = 1$, as we wished to show.

We now move on to the construction of the indicator $\eta_4$. We begin with the proof that the structure of the $U(1)\times\ZT$ symmetry group forces the flux anyon to be time-reversal invariant, $\T(f)=f$, which then guarantees that $f$ can be assigned a well-defined value of $\T^2_f$. Recall that the flux anyon $f$ (which is guaranteed to exist by the theorem that we 
discussed in Sec.~\ref{sec:boson-physical-picture}) satisfies
\beq
	\theta_{f,a}= 2\pi q_a \ (\text{mod }2\pi)\ ,
\eeq
where $\theta_{f,a}$ again denotes the mutual statistics between anyons $f$ and $a$. Under time-reversal the mutual statistics between two anyons $a$ and $b$ satisfies $\theta_{\T(a),\T(b)}= -\theta_{a,b}$. Using this property, we find that
\begin{align}
	\theta_{\T(f),\T(a)} = -2\pi q_a = 2\pi q_{\T(a)}\ ,
\end{align}
where we used Eq.~\eqref{eq:q-U1-times-T} which holds in the $U(1)\times\ZT$ case. By relabeling $b= \T(a)$, we find that
\beq
	\theta_{\T(f),b}= 2\pi q_b
\eeq
for all $b$, which is the same equation satisfied by $f$ itself. It then follows from the unitarity of the $S$-matrix that $\T(f)=f$.

To construct an anomaly indicator that detects whether $f$ is a Kramers singlet or doublet, we again use Eq.~\eqref{eq:NA-formula} for the Kronecker delta function $\delta_{bf}$. We define the anomaly indicator 
\begin{align*}
	\eta_4= \T^2_f = \frac{1}{D}\sum_{a,b\in\C}d_a S_{ab} e^{i2\pi q_a} \T^2_b\ ,
\end{align*}
Note that, by construction, $\eta_4$ can only take the values $\pm 1$, and so the anomaly measured by this indicator
has a $\mathbb{Z}_2$ structure just like the other indicators $\eta_1$, $\eta_2$, and $\eta_3$.

Finally, we mention here that in the $U(1)\times \ZT$ case the two indicators $\eta_3$ and $\eta_4$, which detect whether the flux anyon $f$ is a boson or fermion ($\eta_3$) and a Kramers singlet or doublet ($\eta_4$), correspond exactly to the physical characterization of the ``surface vortex/bulk monopole'' for bosonic SPT phases with $U(1)\times\ZT$ symmetry as discussed in Sec.~4 and Table~1 of Ref.~\onlinecite{senthil-review}.

\subsection{Fermionic case}
\label{sec:U1-times-T-fermions}

We now move on to fermionic systems with $U(1)\times\ZT$ symmetry.
As in the $U(1)\rtimes\mathbb{Z}_2^T$ case, we assume that time-reversal obeys
\begin{align*}
\T^2 = (-1)^Q = P_F
\end{align*}
where $P_F$ is fermion parity and $Q$ is the ``charge" operator, i.e. the generator of the $U(1)$ symmetry. Also, as before we assume that the electron $e$ carries charge $1$ under the $U(1)$ symmetry. (For an example of such a system, consider any electron system with time-reversal symmetry and $S^z$ conservation and define $Q = 2 S^z$). 

Fermionic SPT phases with $U(1)\times\ZT$ symmetry in $(3+1)$-D are believed to have a $\mathbb{Z}_8\times\mathbb{Z}_2$ classification~\cite{wang-senthil,freed-hopkins}. Here the $\mathbb{Z}_8$ factor arises from a reduction of the free fermion classification $\mathbb{Z}\to\mathbb{Z}_8$ in the presence of interactions ($U(1)\times\ZT$ corresponds to symmetry class AIII~\cite{schnyder2008,kitaev2009periodic}). The eight phases corresponding to this $\mathbb{Z}_8$ factor are simply $U(1)$ symmetric versions of the eight ``even'' phases that appear in the $\mathbb{Z}_{16}$ classification of time-reversal symmetric $(3+1)$-D fermionic SPT phases~\cite{wang-senthil,metlitski2014}.
On the other hand, the root phase for the additional $\mathbb{Z}_2$ factor can be thought of as a combination of a bosonic SPT phase --- specifically the SPT phase that supports the ECMC state on its surface --- and a trivial insulating state of electrons.

It follows from these classification results that fermionic topological orders with $U(1)\times\ZT$ symmetry have an anomaly valued in $\mathbb{Z}_8\times\mathbb{Z}_2$. Furthermore, it is clear that the $\mathbb{Z}_8$ factor can be detected by the anomaly indicator $\eta_{2,f}$, discussed in Sec.~\ref{sec:ferm-review}, since the corresponding SPT phases require only time-reversal symmetry. All that remains is to find an anomaly indicator that can detect the missing $\mathbb{Z}_2$ factor in the classification. Here we propose that the indicator $\eta_{3,f}$, defined in Eq.~\eqref{eta3f}, does the job. 

To establish this conjecture, we need to show that $\eta_{3,f}$ has all the desired properties: (1) $\eta_{3,f}$ only takes the values $\pm 1$; (2) $\eta_{3,f} = 1$ for any strictly two-dimensional topological order; and (3) $\eta_{3,f} = -1$ for any topological order that can be realized on the surface of the SPT phase corresponding to the ECMC$+e$ state. The last two properties can be proven starightforwardly using the same reasoning as in Sec.~\ref{sec:eta3feti}; as for property $(1)$, we do not have a general proof but we check it below with several examples.

The examples that we consider are the EfMf$+e$, ETMT$+e$, and ECMC$+e$ states, which we previously discussed, as well as an additional topological order known as the $T_{96}$ state. The $T_{96}$ state is a non-Abelian topological order which contains 96 anyons (including the electron). This topological order was constructed in Ref.~\onlinecite{metlitski2014}, and we review its anyon content and other properties in Appendix~\ref{app:96}. The $T_{96}$ state has been argued to appear on the surface of the root state of the $\mathbb{Z}_8$ part of the classification of fermionic SPTs with $U(1)\times\ZT$ symmetry in $(3+1)$-D~\cite{metlitski2014}. 

In Table~\ref{tab:U1-times-T-fermions} we display the values of the indicators $\eta_{2,f}$ and $\eta_{3,f}$ for the EfMf$+e$, ETMT$+e$, ECMC$+e$, and $T_{96}$ topological orders. We can draw several conclusions from this table. First, the table shows that these two indicators are independent of each other. Second, we can see that these two indicators are able to detect the anomalies in both the ECMC$+e$ and $T_{96}$ topological orders as well as in any combination of the two. Assuming that the $T_{96}$ and ECMC$+e$ states describe the surfaces of the two root phases in the $\mathbb{Z}_8\times\mathbb{Z}_2$ classification, it follows that $\eta_{2,f}$ and $\eta_{3,f}$ are a complete set of anomaly indicators for fermionic topological orders with $U(1)\times\ZT$ symmetry.

As an aside, we note that the anomaly indicator values in Table~\ref{tab:U1-times-T-fermions} show that the ETMT$+e$ state has the same anomalies as four copies of the $T_{96}$ state, which is consistent with the discussion in Sec.~III.B of Ref.~\onlinecite{wang-senthil}. Also, we can see that the EfMf$+e$ state has the same anomalies as a weakly interacting mixture of the ECMC$+e$ and ETMT$+e$ states.

\begin{table}[t]
\begin{center}
\begin{tabular}{ c |c| c| c | c}
 & EfMf$+e$ & ETMT$+e$ & ECMC$+e$ & $T_{96}$ \\ \hline
  $\eta_{2,f}$ & $-1$ & $-1$ & 1 & $ e^{i\frac{\pi}{4}}$ \\ \hline
  $\eta_{3,f}$ & $-1$ & 1 & $-1$ & 1\\
\end{tabular} 
\caption{\label{tab:U1-times-T-fermions} Values of the anomaly indicators $\eta_{2,f}$ and $\eta_{3,f}$ for the
fermionic topological orders EfMf$+e$, ETMT$+e$, ECMC$+e$, and $T_{96}$. The $T_{96}$ and ECMC$+e$
topological orders are two topological orders that can appear on the surfaces of the root phases
for the $\mathbb{Z}_8$ and $\mathbb{Z}_2$ parts, respectively, of the $\mathbb{Z}_8\times\mathbb{Z}_2$
classification of fermionic SPTs with $U(1)\times\ZT$ symmetry in $(3+1)$-D. We have
also included the values of the anomaly indicators for the closely related EfMf$+e$ and ETMT$+e$
topological orders.}
\end{center}
\end{table}

\section{Conclusion}
\label{sec:conclusion}

In this paper we have constructed anomaly indicators for $(2+1)$-D topological orders with time-reversal and $U(1)$ symmetry. These anomaly indicators are functions that take as input the algebraic data describing the symmetric topological order and output a number indicating whether the topological order is anomalous. We have proposed such anomaly indicators for both bosonic and fermionic topological orders with the symmetry groups $U(1)\rtimes\ZT$ and 
$U(1)\times\ZT$. In all cases, we have argued that our indicators are complete in the sense that the anomalies they detect are in one-to-one correspondence with the known
classification of $(3+1)$-D SPT phases with the same symmetry. 

One possible direction for future work would be to find an interpretation of each of our indicators as being equal to the partition function of the corresponding bulk SPT phase on a particular Euclidean four-manifold $X$ and in the presence of a particular background $U(1)$ gauge field $A$. It was already shown in Ref.~\onlinecite{barkeshli2} that $\eta_1$ and $\eta_2$ have a partition function interpretation, and in this paper we have shown in Sec.~\ref{sec:eta-3-pf} that $\eta_3$ also has a partition function interpretation. It would be interesting to find partition function interpretations for the other indicators discussed in this paper.

For the particular indicator $\eta_4$, we can make an educated guess about the correct partition function interpretation. This guess is based on a comparison of our results for $\eta_4$ with the effective actions for SPT phases studied by Kapustin in Ref.~\onlinecite{kapustin1}. Specifically, we conjecture that $\eta_4$ is the partition function for the bulk SPT phase on the manifold $X=\mathbb{RP}^2\times S^2$ and in the presence of a background $U(1)$ gauge field $A_{2\pi}$ that has $2\pi$ flux through the $S^2$ factor, $\int_{S^2} F_{2\pi} = 2\pi$, where $F_{2\pi}=dA_{2\pi}$. We can express this as
\beq
	\eta_4= \mathcal{Z}_{A_{2\pi}}(\mathbb{RP}^2\times S^2)\ , \label{eq:eta-4-pf}
\eeq
where the subscript indicates that we consider the partition function on $\mathbb{RP}^2\times S^2$ and in the presence of $A_{2\pi}$. The evidence for this conjecture is as follows: consider the four topological orders analyzed in Table~\ref{tab:U1-times-T-bosons}. These topological orders are believed to live on the surfaces of the four root phases in the $(\mathbb{Z}_2)^4$ classification of bosonic SPT phases with $U(1)\times\ZT$ symmetry. In Ref.~\onlinecite{kapustin1}, Kapustin proposed effective actions $S_j[X,A]$, $j=1,\dots,4$, for these four root phases, where the notation indicates that $S_j[X,A]$ depends on the data of the manifold $X$ and background gauge field $A$. The partition function for root phase $j$ on $X$ and in the presence of $A$ is then given by $e^{iS_j[X,A]}$. Assuming these effective actions are correct, Eq.~\eqref{eq:eta-4-pf} predicts that the partition function 
$\mathcal{Z}_{A_{2\pi}}(\mathbb{RP}^2\times S^2)$ should give the same result as $\eta_4$ evaluated on the corresponding surface topological order. To check this, note that Table~\ref{tab:U1-times-T-bosons} shows that $\eta_4 = -1$ for the topological order ETMC, and $\eta_4 = 1$ for the other three topological orders. Likewise, one can check that $S_4[\mathbb{RP}^2\times S^2,A_{2\pi}]=\pi$, where $S_4[X,A]= \pi\int_X w_1^2 \wedge \frac{F}{2\pi}$ is the effective action corresponding to ETMC, while $S_j[\mathbb{RP}^2\times S^2,A_{2\pi}]=0$ for $j=1,2,3$. In other words, $\mathcal{Z}_{A_{2\pi}}(\mathbb{RP}^2\times S^2)=-1$ for the root phase that can host ETMC on its boundary, while $\mathcal{Z}_{A_{2\pi}}(\mathbb{RP}^2\times S^2)=1$ for the other three root phases. This matching strongly suggests that Eq.~\eqref{eq:eta-4-pf} is correct, but it would be interesting to find a proof.

\acknowledgements

We thank M. Barkeshli for a discussion about the partition function interpretation of the indicator $\eta_4$, and M. 
Metlitski for a discussion about the $T_{96}$ topological order from Ref.~\onlinecite{metlitski2014}. We thank Seth Musser for pointing out an error in an earlier version of our proof that $\nti = 1$ for Abelian topological orders.
M.F.L. and M.L. acknowledge the support of the Kadanoff Center for Theoretical Physics at the University of Chicago. 
This research was supported in part by NSF DMR-1254741. M.F.L was supported by the Simons Foundation through the ``Ultra-Quantum Matter'' Simons Collaboration.

\appendix

\section{Fermionic topological orders}
\label{app:TO-with-e}

In this appendix we summarize some of the important properties of fermionic topological orders --- topological orders that feature a fermionic excitation $e$ that has trivial mutual statistics with respect to all other anyons.

In general, there are two possible conventions for dealing with the ``electron" $e$: one can either treat $e$ as distinct from the (bosonic) vacuum excitation $1$, or one can identify $e$ with $1$ at the cost of introducing a sign ambiguity in the topological spin. In this paper, we use the first convention: we include $e$ as a distinct quasiparticle in our set of anyons, $\C$. In technical terms, this means that $\C$ is not a unitary modular tensor category since $e$ and $1$ have the same mutual statistics with respect to every other anyon.

The electron $e$ has several special properties. First, for all $a \in \C$ we have $(a\times e)\times e = a$. In other words, fusion with $e$ is an involution in $\C$. Second, since the electron $e$ is a trivial fermion, it follows that for all $a\in\C$,
\begin{align}
	d_{a\times e} = d_{a} \ ; \quad e^{i\theta_{a\times e}} = -e^{i\theta_a} \ .
\end{align}
The final property of $e$ is that the $S$ matrix obeys
\beq
	S_{ae}=S_{e a}= \frac{d_a}{D} \ , \label{eq:locality}
\eeq
for all $a\in\C$~\cite{lan-kong-wen}. Equivalently, the monodromy scalar component (introduced in Eq.~\eqref{eq:MS}) obeys
\beq
	M_{ae}= \frac{1}{d_{e}}=1\ ,
\eeq
Both of these relations are equivalent to the statement that $e$ has trivial mutual statistics with all other anyons: $\theta_{e,a}=1$ for all $a$.

In fermionic topological orders, the $S$ matrix can still be defined via the relation $S_{ab}=\frac{1}{D}\sum_{c\in\C} N^c_{a\ov{b}} d_c e^{i\theta_c}e^{-i\theta_a}e^{-i\theta_b}$, but this $S$ is not unitary due to the presence of the electron. In addition, the matrix $S_{ab}$ defined in this way has the following symmetries under fusion with $e$, 
\beq
	S_{ab}= S_{a\times e,b}= S_{a,b\times e}= S_{a\times e, b\times e}\ . \label{eq:S-symmetries}
\eeq
This symmetry implies that the full $S$ matrix can be written (in a suitable basis) as a tensor product
\beq
	S=\frac{1}{\sqrt{2}}\begin{pmatrix}
	1 & 1 \\
	1 & 1 
	\end{pmatrix}\otimes\hat{S}\ ,
\eeq
where the new matrix $\hat{S}$ is a \emph{unitary} matrix which is half the size of $S$. Note that $\hat{S}$ is unitary because we assume that there are no
other anyons, besides $1$ and $e$, which have trivial mutual statistics with all other anyons. Note also that using this decomposition it is easy to see that the full $S$ matrix is not unitary.

The unitary matrix $\hat{S}$ can be thought of as an $S$ matrix for a theory $\hat{\C}$ which is obtained from $\C$ by taking a quotient in which we identify two anyons if they differ by fusion with $e$, i.e. $b\sim a$ if $b=a\times e$. This partitions $\C$ into equivalence classes $[a]$, each containing two elements: $\{a,a\times e\}$. We denote by $\hat{\C}$ the set whose elements are the equivalence classes $[a]$. The rows and columns of $\hat{S}$ are indexed by these equivalence classes, and we can define the matrix elements of $\hat{S}$ by
\beq
	\hat{S}_{[a][b]}= \sqrt{2}S_{ab}\ .\label{eq:reduced-S}
\eeq
The factor of $\sqrt{2}$ here comes from the ratio between the total quantum dimension $D$ of $\C$ and the total quantum dimension $\hat{D}$ of $\hat{\C}$:
$\hat{D}^2= \sum_{[a]\in\hat{\C}}d_{[a]}^2=\frac{1}{2}D^2$. Note also that there is no ambiguity in this definition of $\hat{S}_{[a][b]}$ since the original $S$ matrix had the symmetries shown in Eq.~\eqref{eq:S-symmetries}. 

\section{Derivation of formula for $\eta_{3,f}$}
\label{app:eta3f}
In this appendix, we derive the formula
\begin{align}
\eta_{3,f} = e^{i\frac{2\pi}{8}(c_- - \sigma_H)},
\label{eta3fapp}
\end{align}
which holds for any $U(1)$ symmetric fermionic topological order $\mathcal{C}$ that can be realized in strictly two dimensions. 

The first step of the proof is to `gauge' the $\mathbb{Z}_2$ fermion parity symmetry in $\mathcal{C}$. This gauging procedure is defined as follows: take a two-dimensional $U(1)$ symmetric lattice model that realizes $\mathcal{C}$, and then minimally couple it to a (weakly-coupled) dynamical $\mathbb{Z}_2$ gauge field in such a way that the microscopic fermions carry $\mathbb{Z}_2$ gauge charge. We will refer to the resulting model as the `gauged' lattice model.

Consider the topological order $\tilde{\mathcal{C}}$ realized by the gauged lattice model. It is not hard to see that $\tilde{\mathcal{C}}$ contains twice as many anyons as $\mathcal{C}$. More specifically, if $\mathcal{C} = \{a,b,c,...\}$, then
\begin{align}
\tilde{C} = \{a, b, c,...\} \cup \{a', b', c',...\}
\end{align}
where $a' = a \times \chi$ and where $\chi$ is the `$\pi$-flux' anyon, i.e. the excitation that is created by adiabatically inserting $\pi$-flux for the $U(1)$ gauge field through some region in the 2D plane. Note that $\chi$ is a real excitation of the gauged lattice model since the dynamical $\mathbb{Z}_2$ gauge field can support $\pi$-fluxes.

To proceed further, observe that $\tilde{\mathcal{C}}$ is a \emph{bosonic} topological order since the only anyon $a \in \tilde{\mathcal{C}}$ that has trivial mutual statistics with respect to every other anyon is the (bosonic) vacuum excitation $1$. (The electron $e$ does not have this property as it has non-trivial mutual statistics with $\chi$: $\theta_{e,\chi} = \pi$). This observation is important because it means we can use the standard (bosonic) formula~\cite{kitaevhoneycomb} to relate the chiral central charge $\tilde{c}_-$ of $\tilde{\mathcal{C}}$ to its anyon content:
\begin{align}
e^{i \frac{2\pi }{8}\tilde{c}_-} = \frac{1}{\tilde{D}} \left(\sum_a d_a^2 e^{i\theta_a} + \sum_{a'} d_{a'}^2 e^{i\theta_{a'}} \right)
\label{cminus1}
\end{align}

The next step is to simplify the right hand side of Eq.~\eqref{cminus1} using properties of the $\pi$-flux anyon $\chi$. These properties mirror the properties of the $2\pi$-flux anyon discussed in Secs.~\ref{sec:boson-physical-picture} and \ref{sec:btideriv}. The first property is that the mutual statistics between $\chi$ and any anyon $a \in \mathcal{C}$ is the Aharonov-Bohm phase
\begin{align}
e^{i\theta_{\chi,a}} = e^{i \pi q_a} 
\end{align}
The second property is that the topological spin/exchange statistics of $\chi$ is
\begin{align}
e^{i\theta_{\chi}} = e^{i \frac{\pi}{4} \sigma_H}
\end{align}
where $\sigma_H$ is the Hall conductivity of the original (ungauged) lattice model. The final property is that $\chi$ is an \emph{Abelian} anyon. 

It follows from these properties of $\chi$ that 
\begin{align}
e^{i \theta_{a'}} &= e^{i (\theta_a + \theta_{\chi,a} + \theta_\chi)} = e^{i (\theta_a + \pi q_a + \pi \sigma_H/4)}
\end{align}
Likewise, we have
\begin{align}
d_{a'} = d_a, \quad \quad \tilde{D} = D\sqrt{2} 
\end{align}
Substituting these relations into Eq.~\eqref{cminus1}, we derive
\begin{align}
e^{i \frac{2\pi }{8}\tilde{c}_-} = \frac{1}{{D}\sqrt{2} } 
\left(\sum_a d_a^2 e^{i\theta_a} + \sum_a d_a^2 e^{i (\theta_a + \pi q_a + \pi \sigma_H/4)} \right)
\label{cminus2}
\end{align}

Next, notice that the first sum on the right hand side of Eq.~\eqref{cminus2} vanishes due to the cancellation between terms involving $a$ and $a \times e$, since $d_a = d_{a \times e}$ and $e^{i\theta_a} = -e^{i\theta_{a\times e}}$. (The second sum does not exhibit this cancellation due to the extra factor of $e^{i\pi q_a}$). Dropping this sum, we derive
\begin{align}
e^{i \frac{2\pi }{8}(\tilde{c}_- - \sigma_H)} = \frac{1}{D\sqrt{2}} 
\sum_a d_a^2 e^{i (\theta_a + \pi q_a)}
\label{cminus3}
\end{align}
To complete the derivation, we note that the gauged lattice model has the same chiral central charge as the original lattice model, i.e. $\tilde{c}_- = c_-$, since the gauging procedure does not introduce new edge modes.\footnote{A more precise way to say this is that it is possible to construct a fully gapped boundary separating the gauged lattice model and the original lattice model.} Making this substitution in (\ref{cminus3}) gives the desired formula (\ref{eta3fapp}).

\section{T-Pfaffian$_{\pm}$ topological orders for the surface of the electron topological insulator (ETI)}
\label{app:TPf}

In this appendix we review the T-Pfaffian$_{\pm}$ topological orders which have been proposed for the surface of the
ETI state in Refs.~\onlinecite{CFV-2014,BNQ}. We compute the anomaly indicators $\eta_{2,f}$, $\eta_{3,f}$, and $\nti$ for these
topological orders in Sec.~\ref{sec:U1-rtimes-T-fermions} of the main text. 
Our discussion of these topological orders follows Ref.~\onlinecite{BNQ}, 
where they were both referred to as ``anyon model X''. The T-Pfaffian$_{\pm}$ topological orders contain 12 anyons, 
denoted by $I_k$, $\psi_k$, for $k=0,2,4,6$, 
and $\sigma_k$ for $k=1,3,5,7$. Here $I,\psi,\sigma$ are Ising anyons (with their usual fusion rules), 
while the $k$ index is an additional $U(1)$ index which takes values in $0,1,\dots,7$. However, the
$k$ index takes only even values for the $I_k$ and $\psi_k$ anyons, and only odd values for the $\sigma_k$ anyons. Thus, 
there are 12 anyons in total, and the total quantum dimension is $D= \sqrt{4+4+(\sqrt{2})^2\times 4}= 4$, where
$\sqrt{2}$ is the quantum dimension of a $\sigma_k$ anyon (the $I_k$ and $\psi_k$ anyons have quantum dimension 
equal to $1$). The vacuum anyon is identified with $I_0$ and the electron is $e =\psi_4$.The $U(1)$ charge of the anyons is determined by their $k$ index only and
is given by 
\beq
	q_{X_k}= -\frac{k}{4}\ \text{mod }2\ ,
\eeq
for $X=I,\psi,\sigma$. All anyons are invariant under
time-reversal \emph{except} for the two pairs $I_2 \leftrightarrow \psi_2$ and $I_6 \leftrightarrow \psi_6$ which are exchanged
by $\T$. Note also that in these two pairs, the anyons \emph{do not} differ by fusion with the electron $\psi_4$. Therefore,
to compute the indicator $\eta_{2,f}$, we only need the $\T^2_a$ values for the eight anyons 
$I_0,\psi_0,I_4,\psi_4,\sigma_1,\sigma_3,\sigma_5$, and $\sigma_7$, which are the anyons that are invariant under 
time-reversal.

The authors of Ref.~\onlinecite{BNQ} found that there are actually two consistent assignments of topological spins and
$\T^2_a$ values for the anyons in their ``anyon model X''. The first possibility (in which their parameter 
$w=-\frac{1}{2}$) corresponds to T-Pfaffian$_{+}$, and the second possibility (in which their parameter 
$w=\frac{7}{2}$) corresponds to T-Pfaffian$_{-}$. The topological spins $e^{i\theta_a}$ and $\T^2_a$ values 
for the anyons in these two cases are shown in Table~\ref{tab:anyon-model-X}. 

\begin{table}[t]
\begin{center}
\begin{tabular}{ c |c| c| c| c|c|c|c|c|c|c|c|c}
  & $I_0$ & $\psi_0$ & $I_2$ & $\psi_2$ & $I_4$ & $\psi_4$  & $I_6$ & $\psi_6$ & $\sigma_1$ & $\sigma_3$ & $\sigma_5$ & $\sigma_7$ \\ \hline 
$e^{i\theta_a}$ ($+$ case) & $1$ & $-1$ & $-i$ & $i$ & $1$ & $-1$ & $-i$ & $i$ & $1$ & $-1$ & $-1$ & $1$  \\ \hline 
$e^{i\theta_a}$ ($-$ case) & $1$ & $-1$ & $i$ & $-i$ & $1$ & $-1$ & $i$ & $-i$ & $-1$ & $1$ & $1$ & $-1$  \\ \hline 
$\T^2_a$ ($+$ case) & $1$ & $1$ &  &  & $-1$ & $-1$ & & & $1$ & $-1$ & $-1$ & $1$ \\ \hline 
$\T^2_a$ ($-$ case)  & $1$ & $1$ &  &  & $-1$ & $-1$ & & & $-1$ & $1$ & $1$ & $-1$  \\ 
\end{tabular} 
\end{center}
\caption{\label{tab:anyon-model-X}Topological spins $e^{i\theta_a}$ and $\T^2_a$ values for the anyons in the 
T-Pfaffian$_{\pm}$ topological orders, which correspond to the cases $w=-\frac{1}{2}$ ($+$ case) and 
$w=\frac{7}{2}$ ($-$ case) of the ``anyon model X'' from Ref.~\protect{\onlinecite{BNQ}}. Note that the $\T^2_a$ entries for 
$I_2,\psi_2,I_6$, and $\psi_6$ are blank because
these anyons are not invariant under time-reversal (they also do not satisfy $\T(a)=a\times e$), and so these anyons do
not possess a well-defined value of $\T^2_a$.}
\end{table}

We conclude this appendix with a few remarks about the $S$ matrix for the T-Pfaffian$_{\pm}$ topological orders, as this matrix is needed for the calculation of the anomaly indicator $\nti$. Actually, it is easier to work with the reduced $S$ matrix $\hat{S}$ for fermionic topological orders which we defined in Appendix~\ref{app:TO-with-e}. 
From Ref.~\onlinecite{BNQ}, a convenient basis for the reduced topological order $\hat{\C}$ is given by
\beq
	\hat{\C}= \{ [I_0], [\psi_0],[\sigma_1], [I_2],[\psi_2],[\sigma_3]\}\ ,
\eeq
where $I_0, \psi_0,\sigma_1, I_2,\psi_2$, and $\sigma_3$ are six anyons in the theory which are \emph{not} related to 
each other by fusion with the electron. The reduced $S$ matrix $\hat{S}$ for $\hat{\C}$ in this basis
is shown explicitly in Eq.~(A.52) of Ref.~\onlinecite{BNQ}, and this $\hat{S}$ is the matrix that we used to calculate
$\nti$ for this topological order.

\begin{table}[t]
\begin{center}
\begin{tabular}{ c |c| c}
 & $e^{i\theta_a}$ & $\td{\T}^2_a$ \\ \hline \hline
$I_{0,0}$ & 1 & 1 \\ \hline
$I_{0,4}$ & 1 & $-1$ \\ \hline
$I_{4,0}$ & 1 & $-1$ \\ \hline
$I_{4,4}$ & 1 & 1 \\ \hline \hline
$\psi_{0,0}$ & $-1$ & 1 \\ \hline
$\psi_{0,4}$ & $-1$ & $-1$ \\ \hline
$\psi_{4,0}$ & $-1$ & $-1$ \\ \hline
$\psi_{4,4}$ & $-1$ & 1 \\ 
\end{tabular} 
\end{center}
\caption{\label{T96-1}Topological spins $e^{i\theta_a}$ and $\td{\T}^2_a$ values for those 
anyons in the set $I_{k,m}$ and $\psi_{k,m}$ which satisfy $\T(a)=a$. For these anyons we have $\T^2_a= \td{\T}^2_a$.}
\end{table}

\section{$T_{96}$ topological order for the surface of a fermionic SPT with $U(1)\times\ZT$ symmetry}
\label{app:96}

In this appendix we review the $T_{96}$ topological order which was derived in Sec.~IV.A of Ref.~\onlinecite{metlitski2014}.
This is a non-Abelian topological order which can appear on the surface of a fermionic SPT phase with $U(1)\times\ZT$ symmetry
(specifically, the root phase of the $\mathbb{Z}_8$ factor in the $\mathbb{Z}_8\times\mathbb{Z}_2$ classification of fermionic
SPTs with $U(1)\times\ZT$ symmetry in $(3+1)$-D). We compute the anomaly indicators $\eta_{2,f}$ and $\eta_{3,f}$ for 
this state in Sec.~\ref{sec:U1-times-T-fermions} of the main text. Below, we summarize the anyon content and basic properties of this
state.

The anyon content of the $T_{96}$ topological order is
\begin{align*}
	I_{k,m} &= I_k e^{im\phi}\ ,\ m=0,\dots,7,\ k=0,2,4,6\ ,\\
	\psi_{k,m} &= \psi_k e^{im\phi}\ ,\ m=0,\dots,7,\ k=0,2,4,6\ , \\
	\sigma_{k,m} &= \sigma_k e^{im\phi}\ ,\ m=0,\dots,7,\ k=1,3,5,7\ .
\end{align*}
Here $I_k,\psi_k,\sigma_k$ are Ising anyons which are labeled by an additional integer
$k\in\{0,1,\dots,7\}$, but with the restriction that $k$ is even for $I_k,\psi_k$ and $k$ is odd 
for $\sigma_k$. Under fusion $I_k,\psi_k,\sigma_k$ obey the usual Ising fusion rules and the extra indices 
$k$ also add modulo $8$. For example, we have $\sigma_k \times \sigma_{k'}= I_{k+k'\text{ mod }8} + \psi_{k+k'\text{ mod }8}$.
Next, the factor $e^{im\phi}$ represents an Abelian anyon labeled by a second integer $m\in\{0,\dots,7\}$. Under fusion, the $m$ indices of these anyons simply add modulo $8$, $e^{im\phi}\times e^{im'\phi}= e^{i\left(m+m'\text{ mod }8\right)\phi}$. As we discuss below, the index $m$ completely determines the charge $q_a$ of each anyon.  

\begin{table}[t]
\begin{center}
\begin{tabular}{ c |c| c}
 & $e^{i\theta_a}$ & $\td{\T}^2_a$ \\ \hline \hline
$I_{2,2}$ & $i$ & $-1$ \\ \hline
$I_{2,6}$ & $i$ & 1 \\ \hline
$I_{6,2}$ & $i$ & 1 \\ \hline
$I_{6,6}$ & $i$ & $-1$ \\ \hline \hline
$\psi_{2,2}$ & $-i$ & $-1$ \\ \hline
$\psi_{2,6}$ & $-i$ & 1 \\ \hline
$\psi_{6,2}$ & $-i$ & 1 \\ \hline
$\psi_{6,6}$ & $-i$ & $-1$ \\
\end{tabular} 
\end{center}
\caption{\label{T96-2}Topological spins $e^{i\theta_a}$ and $\td{\T}^2_a$ values for those 
anyons in the set $I_{k,m}$ and $\psi_{k,m}$ which satisfy $\T(a)=a\times e$. For these anyons we
have $\td{\T}^2_a=-i\T^2_a$.}
\end{table}

There are $96=3\times 4\times 8$ anyons in total in the $T_{96}$ state, and the total quantum dimension is
\beq
	D=\sqrt{32+32+(\sqrt{2})^2 \times 32}= 8\sqrt{2}\ ,
\eeq
where $\sqrt{2}$ is the quantum dimension of a $\sigma_k$ anyon.
The topological spins of these anyons are $e^{i\theta_{I_{k,m}}}$, 
$e^{i\theta_{\psi_{k,m}}}$, and $e^{i\theta_{\sigma_{k,m}}}$ with
\begin{align}
	\theta_{X_{k,m}} &= \theta_X - \frac{\pi k^2}{8} - \frac{2\pi m k}{8} 
\end{align}
where $X=I,\psi,\sigma$ and $\theta_I=0$, $\theta_{\psi}=\pi$, and $\theta_{\sigma}=\frac{\pi}{8}$ as in the usual
Ising anyon theory. Next, the $U(1)$ charge of these anyons depends only on $m$ and is given by
\beq
	q_{X_{k,m}}= \frac{m}{4}\ \pmod{2}\ . \label{eq:T96-anyon-charges}
\eeq
In this topological order the vacuum quasiparticle is identified with $I_{0,0}$ and the electron is 
\beq
	e= \psi_{0,4}\ .
\eeq

To compute the anomaly indicator $\eta_{2,f}$ for the $T_{96}$ state, we need to know (i) the
set of time-reversal invariant anyons $a$ satisfying $\T(a)=a$, and (ii) the set of anyons that
transform under time-reversal as $\T(a)=a\times e$, where $e=\psi_{0,4}$ is the electron in this system.
For the symmetry group $U(1)\times\ZT$, time-reversal negates the $U(1)$ charge. Thus,
time-reversal always sends $m\to -m$ (where $-m$ is equivalent mod $8$ to one of the 
values $0,\dots,7$). On the other hand, the $k$ index of the anyons in the $T_{96}$ state is invariant under time-reversal. 
To identify the anyons that transform as $\T(a)=a$ and $\T(a)=a\times e$ we must also use the general constraints which
follow from the $U(1)\times\ZT$ group structure, namely that $e^{i\theta_a}= e^{-i\theta_{\T(a)}}$ and 
$q_a=-q_{\T(a)} \pmod{2}$. 

\begin{table}[t]
\begin{center}
\begin{tabular}{ c |c| c  }
 & $e^{i\theta_a}$ & $\td{\T}^2_a$ \\ \hline \hline
$\sigma_{1,0}$ & 1 &  $1$  \\ \hline
$\sigma_{3,0}$ & $-1$ &  $-1$ \\ \hline
$\sigma_{5,0}$ & $-1$ &  $-1$ \\ \hline
$\sigma_{7,0}$ & 1 &  $1$ \\ \hline\hline
$\sigma_{1,4}$ & $-1$ & $-1$  \\ \hline
$\sigma_{3,4}$ & 1 &  $1$  \\ \hline
$\sigma_{5,4}$ & 1 &  $1$  \\ \hline
$\sigma_{7,4}$ & $-1$ & $-1$  \\ 
\end{tabular} 
\end{center}
\caption{\label{T96-3}Topological spins $e^{i\theta_a}$ and $\td{\T}^2_a$ values for the anyons
$\sigma_{k,0}$ and $\sigma_{k,4}$, $k=1,3,5,7$. These anyons satisfy $\T(a)=a$ and so we have
$\T^2_a=\td{\T}^2_a$ for all of these anyons.}
\end{table}

The considerations in the previous paragraph imply that the time-reversal invariant anyons in the $T_{96}$ state must have $m=0$
or $4$ and topological spin $e^{i\theta_a}=\pm 1$, while anyons $a$ that transform as $\T(a)=a\times e$ must
have $m=2$ or $6$ and topological spin $e^{i\theta_a}= \pm i$. Using these conditions one can show that the time-reversal invariant 
anyons are
\begin{align}
	& \{I_{k,0}\ ,\ I_{k,4}\ ,\ \psi_{k,0}\ ,\ \psi_{k,4}\}\ ,  \quad \quad k=0,4\ ; \nonumber \\
	& \{\sigma_{k,0}\ ,\ \sigma_{k,4} \}\ , \quad \quad \quad \quad \quad \quad \quad k=1,3,5,7\ . 
\end{align}
Next, the anyons that transform as $\T(a)=a\times e$ are the
pairs $I_{2,2} \leftrightarrow \psi_{2,6}$, $I_{2,6} \leftrightarrow \psi_{2,2}$, 
$I_{6,2} \leftrightarrow \psi_{6,6}$, $I_{6,6} \leftrightarrow \psi_{6,2}$, and
also $\sigma_{k,2}\leftrightarrow\sigma_{k,6}$ for $k=1,3,5,7$. Here the
double arrow means that the two anyons transform into each other under
time-reversal and that the two anyons differ by fusion with the electron $e$. These can
all be deduced using the fusion rules for the Ising anyons and the fact that the $k$ and $m$ indices simply add under fusion,
for example $I_{2,2}\times e= I_{2,2}\times\psi_{0,4}= \psi_{2,6}$ and
$\sigma_{k,2}\times e= \sigma_{k,2}\times\psi_{0,4}= \sigma_{k,6}$.

In Tables~\ref{T96-1}, \ref{T96-2}, \ref{T96-3}, and \ref{T96-4}, we list the topological spins
$e^{i\theta_a}$ and $\td{\T}^2_a$ values for all anyons $a$ in the $T_{96}$ state that satisfy $\T(a)=a$ or $\T(a)=a\times e$.
These values (along with the charges of the anyons from
Eq.~\eqref{eq:T96-anyon-charges}) are then used in Sec.~\ref{sec:U1-times-T-fermions} 
of the main text to compute the anomaly indicators $\eta_{2,f}$ and $\eta_{3,f}$ for this topological order.

We close this appendix with a brief explanation of how the 
$\td{\T}^2_a$ values shown in Tables~\ref{T96-1}, \ref{T96-2}, \ref{T96-3}, and \ref{T96-4} can be computed using the
data in Ref.~\onlinecite{metlitski2014}. In Ref.~\onlinecite{metlitski2014}, the authors constructed the $T_{96}$ state by
first constructing an intermediate superfluid state (in which the $U(1)$ symmetry of the system is spontaneously broken), 
and then condensing vortices in that superfluid state to restore the $U(1)$ symmetry and obtain the symmetric and topologically 
ordered $T_{96}$ state. The vortices in the superfluid state can be identified with the $X_k$ anyons ($X=I,\psi,\sigma$), i.e., with the 
charge neutral part of the full $X_{k,m}$ anyon in the $T_{96}$ state.\footnote{We follow Ref.~\onlinecite{metlitski2014} and use 
the same notation $X_k$ to denote the neutral part of an $X_{k,m}$ anyon in the $T_{96}$ state and the corresponding vortex in the 
superfluid state.} In particular, if we drive a transition from the $T_{96}$ state 
back to the superfluid state, then each anyon $X_{k,m}$ loses its $U(1)$ charge, which was determined by
the index $m$, and reverts back to its neutral vortex part $X_k$. One very important property of the vortices $X_k$ in the superfluid,
which will be used below, is that the vortex $X_k$ binds $k$ Majorana fermions in its core.

\begin{table}[t]
\begin{center}
\begin{tabular}{ c |c| c}
 & $e^{i\theta}$ & $\td{\T}^2_a$ \\ \hline \hline
$\sigma_{1,2}$ & $-i$ &  $-1$   \\ \hline
$\sigma_{3,2}$ & $-i$ &  $-1$  \\ \hline
$\sigma_{5,2}$ & $i$ &  $1$  \\ \hline
$\sigma_{7,2}$ & $i$ &  $1$ \\ \hline\hline
$\sigma_{1,6}$ & $i$ &  $1$  \\ \hline
$\sigma_{3,6}$ & $i$ &  $1$ \\ \hline
$\sigma_{5,6}$ & $-i$ & $-1$  \\ \hline
$\sigma_{7,6}$ & $-i$ & $-1$ \\ 
\end{tabular} 
\end{center}
\caption{\label{T96-4}Topological spins $e^{i\theta_a}$ and $\td{\T}^2_a$ values for the anyons
$\sigma_{k,2}$ and $\sigma_{k,6}$, $k=1,3,5,7$. These anyons satisfy
$\T(a)=a\times e$, and so we have $\td{\T}^2_a=-i\T^2_a$ for these anyons.}
\end{table}

The intermediate superfluid phase possesses a modified time-reversal symmetry, denoted by $\mathcal{S}$, which is related
to the original time-reversal operation $\T$ by
\beq
	\mathcal{S}= U(\tfrac{\pi}{2})\T\ , \label{eq:S-and-T}
\eeq
where $U(\tfrac{\pi}{2})$ is the unitary operator which implements the $U(1)$ rotation by the angle $\frac{\pi}{2}$. The $k$ index
of the vortices $X_k$ in the superfluid is invariant under the action of $\mathcal{S}$, and so it is possible to study the
transformation properties of the internal states of each vortex under the action of $\mathcal{S}$ 
(recall that each vortex contains a certain number of Majorana fermions
in its core).  In particular, vortices can be assigned a definite value of 
$\mathcal{S}^2$, and this value can be computed using the techniques outlined in Sec.~V of \cite{metlitski2014}.
Using the relation \eqref{eq:S-and-T} between $\mathcal{S}$ and $\mathcal{T}$, the $\T^2$ values for each anyon can then
be obtained as
\beq
	\T^2_a= e^{-i\pi q_a}\mathcal{S}^2_a\ , \label{eq:S2-and-T2}
\eeq
where for an anyon $a=X_{k,m}= X_k e^{i m\phi}$, $\mathcal{S}^2_a$ is equal to the value of $\mathcal{S}^2$ when acting on the
vortex $X_k$ in the intermediate superfluid state. 

Finally, there is one subtle point in the assignment of $\mathcal{S}^2$ values to vortices that we wish to highlight here. 
The subtlety is that when computing the $\T^2_a$ values for an anyon $a$ using Eq.~\eqref{eq:S2-and-T2},  
we must distinguish between bosonic and fermionic transformations of vortices under the action of $\mathcal{S}$. By a bosonic vs. 
fermionic transformation we mean the following. Let $|v\ran$ be a quantum state of the superfluid containing two widely separated vortex 
excitations of the type $X_k$. Then we say that $X_k$ transforms in a bosonic manner under $\mathcal{S}$ if
\beq
	\mathcal{S}|v\ran= b_1 b_2 |v\ran\ ,
\eeq 
where $b_1$ and $b_2$ are \emph{bosonic} operators localized near the first and second $X_k$ vortex~\cite{levin-stern}. Similarly, we 
say that $X_k$ transforms in a fermionic manner under $\mathcal{S}$ if
\beq
	\mathcal{S}|v\ran= c_1 c_2 |v\ran\ ,
\eeq 
where $c_1$ and $c_2$ are \emph{fermionic} operators localized near the first and second $X_k$ vortex~\cite{metlitski2014}.

It turns out that, depending on the value of $k$, we have the following three possibilities for how a vortex $X_k$ can transform under
$\mathcal{S}$: (i) $X_k$ transforms in a bosonic manner under $\mathcal{S}$, (ii) $X_k$ transforms in a fermionic manner under 
$\mathcal{S}$, or (iii) $X_k$ can transform either in a bosonic or a fermionic manner under $\mathcal{S}$. The third possibility arises
due to the fact that for a given state $|v\ran$ it may be possible to find operators $b_1,b_2,c_1$, and $c_2$ such that
\beq
	b_1 b_2 |v\ran= c_1 c_2 |v\ran\ .
\eeq

\begin{table}[t]
\begin{center}
\begin{tabular}{ c|c|c|c|c|c|c|c|c}
$k$ & 0&1&2&3&4&5&6&7 \\ \hline 
$\mathcal{S}^2$ (bosonic) & 1 & 1 & & $-1$ & $-1$ & $-1$ & & 1 \\ \hline
$\mathcal{S}^2$ (fermionic) & & 1 & 1 & 1 & & $-1$ & $-1$ & $-1$ \\
\end{tabular} 
\end{center}
\caption{\label{S2-values} Bosonic and fermionic $\mathcal{S}^2$ values for vortices $X_k$ in the intermediate superfluid phase used to construct the $T_{96}$ state in Ref.~\protect{\onlinecite{metlitski2014}}. If an entry in the table is left blank then the vortex with the specified $k$ value does not transform in a bosonic/fermionic manner under $\mathcal{S}$.}
\end{table}

Now that we understand the different possibilities for how a vortex can transform under $\mathcal{S}$, we are ready to state the
rule that must be used to compute $\T^2_a$ from $\mathcal{S}^2_a$ using Eq.~\eqref{eq:S2-and-T2}: if $a$ in the $T_{96}$ 
topological order transforms under $\T$ as $\T(a)=a$, then we must use the value of $\mathcal{S}^2_a$ that is obtained from the
\emph{bosonic} action of $\mathcal{S}$ on the corresponding vortex in the superfluid. On the other hand, if $\T(a)=a\times e$, then we 
must use the value of $\mathcal{S}^2_a$ that is obtained from the \emph{fermionic} action of $\mathcal{S}$ on the corresponding 
vortex in the superfluid. In Table~\ref{S2-values} we summarize the bosonic and fermionic $\mathcal{S}^2$ values that are needed to compute $\T^2_a$ according to these rules.

%\bibliography{anomaly-indicators-U1-refs}

%merlin.mbs apsrev4-1.bst 2010-07-25 4.21a (PWD, AO, DPC) hacked
%Control: key (0)
%Control: author (8) initials jnrlst
%Control: editor formatted (1) identically to author
%Control: production of article title (-1) disabled
%Control: page (0) single
%Control: year (1) truncated
%Control: production of eprint (0) enabled
%

\end{document}